\documentclass[a4paper,11pt]{article} 
\usepackage{jheppub}
\usepackage{color}
\usepackage{amsmath,amssymb}
\usepackage{hhline}
\usepackage{multirow}
\usepackage{subfigure}
\usepackage{epstopdf}
\usepackage{xcolor}
\usepackage{capt-of}
\usepackage{ulem}
\usepackage{cancel}
\newcommand{\ba}{\begin{eqnarray}}
\newcommand{\ea}{\end{eqnarray}}
\newcommand{\be}{\begin{equation}}
\newcommand{\ee}{\end{equation}}
\newcommand{\bal}{\begin{align}}
\newcommand{\eal}{\end{align}}

\newcommand{\barr}{\begin{eqnarray}}
\makeatletter
\def\fmslash{\@ifnextchar[{\fmsl@sh}{\fmsl@sh[0mu]}}
\def\fmsl@sh[#1]#2{%
  \mathchoice
    {\@fmsl@sh\displaystyle{#1}{#2}}%
    {\@fmsl@sh\textstyle{#1}{#2}}%
    {\@fmsl@sh\scriptstyle{#1}{#2}}%
    {\@fmsl@sh\scriptscriptstyle{#1}{#2}}}
\def\@fmsl@sh#1#2#3{\m@th\ooalign{$\hfil#1\mkern#2/\hfil$\crcr$#1#3$}}
\makeatother

\newcommand{\GeV}{\mbox{GeV}}
\newcommand{\MeV}{\mbox{MeV}}

\definecolor{darkgreen}{RGB}{90,150,50}
\definecolor{brown}{RGB}{150,50,0}

\title{\boldmath
 The $D^*D \pi$ and $B^*B\pi$ couplings
\\ from light-cone sum rules }

\author[a]{Alexander Khodjamirian,}
\author[b]{Bla\v zenka  Meli\'c,}
\author[c]{Yu-Ming Wang}
\author[c,d]{and Yan-Bing Wei}
\affiliation[a]{Theoretische Physik 1, Naturwissenschaftlich-Technische Fakult\"at, Universit\"at Siegen, D-57068 Siegen, Germany}
\affiliation[b]{Rudjer Boskovic Institute, Division of Theoretical Physics, Bijeni\v cka 54, HR-10000 Zagreb, \\Croatia}
\affiliation[c]{School of Physics, Nankai University, Weijin Road 94, 300071 Tianjin, China}
\affiliation[d]{Physik Department T31, James-Franck-Stra\ss e~1, Technische Universit\"at M\"unchen, D-85748 \\
Garching, Germany}

\emailAdd{khodjamirian@physik.uni-siegen.de,
melic@irb.hr,
wangyuming@nankai.edu.cn,
weiyb@nankai.edu.cn}

\preprint{SI-HEP-2020-29, P3H-20-071, RBI-ThPhys-2020-13, TUM-HEP-1297/20
}

\abstract{We revisit the
calculation of the strong couplings $D^*D\pi$ and $B^*B\pi$
from the QCD light-cone sum rules using
the pion light-cone distribution amplitudes.
The accuracy of the correlation function, calculated from the operator
product expansion near the light-cone, is upgraded by
taking into account the gluon radiative corrections to the twist-3 terms. The double spectral
density of the correlation function,
including the  twist-2, 3  terms at ${\cal O} (\alpha_s)$
and the twist-4 LO terms, is presented in an analytical form
for the first time. This form allows  us to
use various versions of the quark-hadron duality regions in the  double
dispersion relation underlying the sum rules.
We predict $g_{D^*D\pi}=14.1^{+1.3}_{-1.2}$ and $g_{B^*B\pi}=30.0^{+2.6}_{-2.4}$
when the decay constants of heavy mesons entering the light-cone sum rule
are taken from lattice QCD results. We
compare our results with the experimental value for the
charmed meson coupling and with the lattice QCD calculations.
\\
\\
}

\begin{document}

\maketitle
\flushbottom

\section{Introduction}
The strong couplings of heavy-light pseudoscalar and vector mesons with the pion
belong to the most important hadronic parameters of heavy flavour physics.
Our ability to calculate these couplings reflects the currently achieved progress
in QCD and related effective theories.
In the charm sector, the $D^*D\pi$  coupling has been measured,
combining the branching fractions of the $D^* \to D \pi$ decays
with the total width of $D^*$.
The latter  is  currently available from
the two experiments \cite{CLEODstDpi,BaBarDstDpi,PDG}  and has a small error.
The $B^*B\pi$ coupling cannot be directly measured, due to the lack of
phase space for a $B^* \to B\pi$ decay. Still, this coupling is phenomenologically very important.
It enters the residue of the $B^*$-meson  pole  in the vector $B\to \pi$ form factor
used for the determination of the CKM parameter $V_{ub}$.
Located very close to the kinematical threshold of the
 $B\to \pi\ell \bar{\nu}_\ell$ semileptonic transitions,
the $B^*$ pole  significantly influences the form factor at small
hadronic recoil.

In the infinitely heavy-quark limit $m_b\to \infty$, the $B^*B\pi$ coupling
turns into the ``static'' strong coupling of heavy-light mesons with the pion,
a key  parameter  in the heavy-meson chiral perturbation theory (HM$\chi$ PT)
\cite{Burdman:1992gh,HMchPT,Yan:1992gz}.
There are several lattice QCD calculations of the heavy-meson strong  couplings
and their static limit, the most advanced ones,
calculated with dynamical quarks, are  in
\cite{Becirevic:2009yb,Becirevic:2012pf,Can:2012tx,Detmold:2012ge,Bernardoni:2014kla,Flynn:2015xna}.

In \cite{BBKR}, the $D^*D\pi$ and $B^*B\pi$ couplings have been calculated,
employing the method of light-cone sum rules (LCSRs) in QCD \cite{lcsr1,lcsr2,lcsr3}.
The extension of LCSRs to strong couplings goes back to \cite{Braun:1988qv}
where the pion-nucleon and $\rho \omega \pi$ couplings were calculated.
The underlying object in this method is the vacuum-to-pion correlation function calculated near the light cone
in terms of the operator product expansion (OPE) involving the universal pion light-cone distribution
amplitudes (DAs) of growing twist.
The same correlation function is used in the well established LCSRs for the $B\to \pi$  and $D\to\pi$ form factors,
see e.g., \cite{BBKR,KRWY1,Bagan:1997bp,BZ04,DKMMO,KKMO}.
Importantly, the calculation of the $D^*D\pi$ and $B^*B\pi$ couplings is performed
at a finite heavy-quark mass. Hence, not only
the infinitely heavy quark limit of these couplings can be taken, but also the inverse mass corrections are accessible.
The LCSR is obtained, employing analyticity in the two external
momenta squared and matching the resulting double
dispersion relation to the OPE result. The further steps follow the  standard QCD sum rule technique and involve the
quark-hadron  duality approximation and the double Borel transformation.
Due to the approximate degeneracy of vector and pseudoscalar heavy-light mesons
(becoming exact  in the infinitely heavy quark limit),
equal  Borel parameters are taken in both channels
of the double dispersion relation.
As a result, the LCSR predictions \cite{BBKR} for the $D^* D\pi$
and $B^*B\pi$ strong couplings at the leading order (LO) in $\alpha_s$
are sensitive to the  values of the pion DAs at
$u=\bar{u} = 1/2$ where $u$ and $\bar{u}\equiv 1-u$ are
the fractions of the pion momentum carried by the collinear
quark and antiquark in the two-parton state of the pion.
The shape of the pion twist-2  DA is usually
described by an expansion in Gegenbauer polynomials based on the conformal
partial-wave expansion.
The value of this  DA at the middle point provides  a nontrivial
constraint on the polynomial coefficients (Gegenbauer moments).
Thus, LCSR for the strong coupling complements the information
on the first few Gegenbauer moments available from other sources
(e.g., lattice QCD calculation of the second moment and LCSRs for the pion form factors).
Assessing the accuracy of the LCSR for the heavy-light strong couplings,
one has to mention that  the use of the double dispersion relation makes this sum rule
more sensitive to the quark-hadron duality approximation than the LCSR for
heavy-to-light form factors based on the single-variable dispersion relation.
On the other hand, the accuracy of OPE in both sum rules is the same.

The interval for $g_{D^*D\pi}$ obtained in \cite{BBKR}
 appeared to be below the measured value by about 30\%.
The heavy-quark limit of this   coupling obtained from LCSR was also
smaller than the  results of lattice QCD calculations.
The gluon radiative correction to the twist-2 term of LCSR
calculated  in \cite{KRWY_2} did not remove this discrepancy.
However, one should mention that theoretical uncertainties quoted
in the previous analyses \cite{BBKR,KRWY_2} are incomplete
and include only a part of parametrical uncertainties.
In particular, the perturbative correction to the twist-3 term
was not taken into account.
The dependence on the form of duality region was also not completely investigated,
moreover, the subleading twist-3, 4 contributions were included without
the duality subtraction at tree level.
Another critical point is the choice of decay constants of pseudoscalar
and vector heavy-light mesons which multiply the strong coupling in LCSR,
making the final result  very sensitive to the values of  these hadronic parameters.

A  possibility to explain the deficit of the LCSR prediction
for the heavy-light strong coupling is to allow for
large contributions of excited heavy-light states to the double dispersion
relation, as pointed out in \cite{Khodjamirian:2001bj}
and discussed in more detail in  \cite{Becirevic:2002vp}. Note however that this conjecture
introduces an almost uncontrollable model-dependence in the hadronic part
of the sum rule and leaves open the most important question: is there a duality
region which effectively corresponds only to the ground-state contribution to the LCSR?
Taking into account all above mentioned open aspects,
it is timely  to revisit the LCSR calculation
of the strong couplings $B^*B \pi$ and $D^*D\pi$, upgrading and
updating the earlier analyses in \cite{BBKR,KRWY_2}.

In this paper we pursue three main goals.
The first one is to improve the accuracy of the OPE for the
underlying correlation function. To this end, we will include
the next-to-leading-order (NLO) twist-3 term,
calculating the corresponding gluon radiative corrections.
We remind that in the LCSRs for the strong couplings
the twist-3 part is comparable to the twist-2 part, their ratio being
of $O(\mu_\pi/m_Q)$, where the chirally enhanced parameter
$\mu_\pi=m_\pi^2/(m_u+m_d)$ is comparable with the heavy quark mass $m_Q=m_{c,b}$.
Hence, by adding the gluon radiative correction to the twist-3 term, we will achieve the
same NLO accuracy for both equally important parts of the OPE. Furthermore, we will,
for the first time, represent both NLO corrections in a form of double dispersion relation
with compact analytical expressions for the double spectral density.

Due to the importance of the twist-3 part, the LCSR
considered here involves a double hierarchy of even (2, 4, 6,..)
and odd (3, 5,..) twist terms. The twist-4 contributions
known from previous analyses will be added in LO, which is sufficiently
accurate since the twist-4 part is small with respect to the twist-2 LO part.
Moreover, the twist-5, 6 contributions to the
underlying correlation function calculated recently \cite{Rusov:2017chr}
in the factorizable approximation were found negligible. Also the next-to-next-to-leading-order (NNLO)
correction to the twist-2 part obtained in \cite{Bharucha:2012wy}
(in the large $\beta_0$ approximation) is very small.
All this ensures that the OPE adopted here, including
the twist-2, 3 terms at NLO, and the twist-4 term at LO, is sufficiently accurate.

Our second goal in this work is to update the input parameters in LCSR.
In particular, in this paper we employ the $\overline{\rm MS}$ mass scheme
for the highly off-shell heavy quarks in the correlation function, which
is a more appropriate choice than  the pole-mass scheme
employed in the earlier calculation.
We also use the latest knowledge on the input parameters of pion DA's.
For the  decay constants of the vector and
pseudoscalar heavy-light mesons we use the QCD two-point sum rules
with the same NLO accuracy as LCSRs, employing  the results of the
updated analysis in \cite{GKPR} as well as the recent lattice QCD results.
A more complete analysis of parametrical uncertainties
of the sum rule results is done.

Finally, our third goal is to extend the quark-hadron duality
approximation for the continuum subtraction to all twist-3 and 4 terms,
in order to improve  the procedure of subtraction of excited states in LCSR
which was incomplete in \cite{BBKR}.
The sensitivity of LCSRs to the form of the quark-hadron duality region
in the double dispersion relation will be investigated.

The plan of this paper is as follows. After outlining  the LCSR method
in Sect.~2,  we present in Sect.~3 the double spectral density of the correlation function
in updated form, including the new twist-3 radiative correction.
In Sect.~4  we discuss different forms of
the quark-hadron duality ansatz for the double dispersion relation.
Sect.~5 contains the numerical analysis and
Sect.~6 is devoted to the concluding discussion.
We present in Appendix~\ref{app:lcda} necessary details on the pion DAs,
in Appendix~\ref{app:nlo} the expressions for the double spectral densities at NLO
and in Appendix {\ref{app:2pt}} the sum rules for the heavy-light meson decay constants.

\section {The LCSR method}

Hereafter we use a generic notation $H^{(*)}$ for both pseudoscalar (vector) mesons
$D^{(*)}$ and $B^{(*)}$.
The strong $H^*H\pi$ coupling $g_{H^*\!H\pi}$ is defined as
the invariant constant parametrizing the hadronic matrix element
\be
\langle H^*(q) \pi(p)|H(p+q)\rangle =-g_{H^*\!H\pi}\,p^\mu\epsilon_\mu^{(H^*)}\,,
\label{eq:strong}
\ee
where the vector and pseudoscalar meson have four-momenta $q$  and $p+q$, respectively, and
$\epsilon_\mu^{(H^*)}$ is the polarization vector of $H^*$.  The infinitely
heavy  quark limit of  the strong coupling:
\be
\lim_{m_Q\to \infty} g_{H^*\!H\pi}/(2m_H)= \hat{g}/f_\pi\,,
\label{eq:heavy}
\ee
where $f_\pi$ is the pion decay constant, determines the static coupling $\hat{g}$
that does not depend on the heavy mass scale
and enters the HM$\chi$PT Lagrangian.

In \cite{BBKR} it was suggested  to calculate the strong couplings
($\ref{eq:strong}$) employing the LCSR based on the light-cone OPE
for the vacuum-to-pion correlation function:
\be
F_\mu(q,p)=i\!\int \! d^4x e^{iqx}\langle \pi(p)|T\{j_\mu(x), j_5(0)\} |0\rangle
= F(q^2,(p+q)^2) \, p_\mu + \dots\,,
\label{eq:corr}
\ee
where $j_\mu=\bar{q}_1\gamma_\mu Q $ and  $j_5=(m_Q+m_{q_2})\bar{Q}\,i\gamma_5 q_2 $
are the interpolating currents for the $H^*$ and $H$ mesons, respectively.
In the above, $Q$ is a generic notation for the heavy quarks $c$ and $b$,
and $q_{1,2}$ stand for the light quarks $u$ or $d$.
The decay constants of heavy-light mesons  needed here are defined as:
\be
\langle 0|j_\mu|H^*(q) \rangle = m_{H^*}\epsilon^{(H^*)}_\mu f_{H^*} , ~~
\langle 0|j_5|H(p+q) \rangle = m_{H}^2 f_{H} \,.
\label{eq:fBst}
\ee
In (\ref{eq:corr}) the relevant invariant amplitude $F$ multiplying $p_\mu$
is singled out, and the second Lorentz structure proportional
to $q_\mu$ is indicated by ellipses. The pion is on shell and in what
follows we adopt the chiral symmetry, putting $p^2=m_\pi^2=0$
and neglecting the $u,d$ quark masses in the correlation function,
adopting also the isospin symmetry.
Note that the enhanced parameter
\be
\mu_\pi=\frac{m_\pi^2}{m_u+m_d}
\label{eq:mupi}
\ee
is retained
in the chiral limit, since $m_\pi^2\sim O(m_u+m_d)$.
For the finite heavy quark mass in (\ref{eq:corr}) we employ the $\overline{\rm MS}$ scheme.

To  derive LCSR for the strong coupling, following \cite{BBKR}
one inserts the complete set of intermediate states with $H$ and $H^*$ quantum
numbers in (\ref{eq:corr}) and employs the double dispersion
relation
\footnote{Double dispersion relations were used for
QCD sum rules based on the local OPE, starting from
\cite{AK80} where the sum rules for charmonium radiative transitions were obtained.
Another important application of double sum rules
is the pion form factor \cite{IoffeSmilga,Nesterenko:1982gc}; for the others
see, e.g., the review \cite{CK}. The first application of LCSRs
for hadronic couplings are presented in \cite{lcsr1,lcsr2,lcsr3,Braun:1988qv}.
}
for the amplitude $F(q^2,(p+q)^2)$ in the two independent
variables $q^2$ and $(p+q)^2$:
\ba
F(q^2,(p+q)^2)&=& \frac{m_H^2 m_{H^*} f_H f_{H^*}g_{H^*H\pi}}
{(m_H^2-(p+q)^2)(m_{H^*}^2-q^2)}
\nonumber\\
&+&\iint\limits_{\!\Sigma} ds_2 ds_1 \frac{\rho^h(s_1,s_2)}{(s_2-(p+q)^2)(s_1-q^2)} + \dots \,,
\label{eq:double_disp}
\ea
where the possible subtraction terms are not shown.
The latter include, in general, single dispersion
integrals in the first variable $(p+q)^2$ combined with polynomials in the second variable $q^2$
and vice versa. All subtraction terms vanish after the double Borel transformation which will  be applied to the relation (\ref{eq:double_disp}).

The ground-state double-pole term in the above relation
contains the product of $H^*H\pi$ strong coupling and decay constants.
We denote by $\Sigma$ the two-dimensional region with the lower boundary
$\{s_1\geq (m_H+m_\pi)^2; \, s_2\geq (m_{H^*}+m_\pi)^2 \}$, where the hadronic
spectral  density of the continuum and excited states  (with the $H^*$
and $H$ quantum numbers, respectively) denoted as $\rho^h(s_1,s_2)$  contributes.

At $q^2, (p+q)^2 \ll m_Q^2$, the dispersion relation (\ref{eq:double_disp})
is matched to the result of the QCD calculation of $F(q^2,(p+q)^2)$.
For the latter, we use the light-cone OPE in terms of pion DAs,
and employ the most complete and up-to-date calculation in \cite{DKMMO}
that was used to obtain the LCSR for the $B\to \pi$ form factor. (see also
\cite{BZ04}, where, however the complete analytical expressions are not presented).
Following the general outline of QCD sum rule  derivation
\cite{Shifman:1978bx}, we employ the quark-hadron duality ansatz.
To this end, we will represent the OPE result for the correlation function
in a form of double dispersion integral:
\begin{align}
F^{\rm (OPE)}(q^2,(p+q)^2)=
\int\limits^\infty_{-\infty} \frac{ds_2}{s_2-(p+q)^2}
\int\limits^\infty_{-\infty}\frac{ds_1}{s_1-q^2}\,\,
\rho^{\rm (OPE)}(s_1, s_2)\,,
\label{eq:ddispOPE}
\end{align}
with the double spectral density
\be
\rho^{\rm (OPE)} (s_1, s_2) \equiv \frac{1}{\pi^2}
\mbox{Im}_{s_1}\mbox{Im}_{s_2} F^{\rm (OPE)}(s_1,s_2)\,,
\label{eq:rhoope}
\ee
to be derived in the next section.
Hereafter, we denote the variables $q^2$ and
$(p+q)^2$ continued to their  timelike regions as $s_1$ and
$s_2$, respectively. For the sake of compactness, the lower limits of integration  in (\ref{eq:ddispOPE}) corresponding to the thresholds $s_{1,2}=m_Q^2$ are formally included
in the spectral densities in a form of step functions and their derivatives. We also omit in
(\ref{eq:ddispOPE}) all subtraction terms  that vanish after double Borel transformation.

Adopting the quark-hadron duality, we assume that
the  integral of the hadronic spectral density $\rho^h(s_1,s_2) $
taken over the two-dimensional region $\Sigma$ in (\ref{eq:double_disp})
is equal to the
integral of the OPE spectral density (\ref{eq:rhoope})
taken over a certain region $\Sigma_0$ in the $(s_1,s_2)$ plane
\begin{align}
\iint\limits_{\!\Sigma}ds_2\,ds_1
\frac{\rho^{h}(s_1, s_2)}{(s_2-(p+q)^2)(s_1-q^2)}=
\iint\limits_{\!\Sigma_0}ds_2\,ds_1
\frac{\rho^{\rm (OPE)}(s_1, s_2)}{(s_2-(p+q)^2)(s_1-q^2)}
\,.
\label{eq:dualapp}
\end{align}

To proceed, we equate the  double dispersion
representations  (\ref{eq:double_disp})  and (\ref{eq:ddispOPE}),
substitute  (\ref{eq:dualapp}) to (\ref{eq:double_disp})
and subtract the equal integrals over the region $\Sigma_0$ from both sides
of this equation. For the remaining region dual to the
ground-state contribution of the $H^*\to H \pi$ transition to (\ref{eq:double_disp})
we introduce a generic notation:
\begin{align}
\iint\limits^{\Sigma_0}ds_2\,ds_1...=
\int\limits^\infty_{-\infty} \! ds_2
\int\limits^\infty_{-\infty}\!ds_1... - \iint\limits_{\!\Sigma_0}ds_2\,ds_1... \,.
\label{eq:dualdef}
\end{align}
The actual choice of this duality region will be discussed below.
As a next step, we perform the double Borel transformation, defined as
\begin{eqnarray}
f(M_1^2,M_2^2) &=& \left [\lim_{\{-q^2,\,n\}\to \infty,  \atop -q^2/n=M_1^2}
\frac{(-q^2)^{n+1}}{n!}\left(\frac{d}{dq^2}\right)^n \right ]
\nonumber\\
&& \left [ \lim_{\{-(p+q)^2,\,k\}\to \infty, \atop  -(p+q)^2/k=M_2^2}
\frac{(-(p+q)^2)^{k+1}}{k!}\left(\frac{d}{d(p+q)^2}\right)^k \right ]
\, f(q^2,(p+q)^2)\,.
\label{eq:Borel}
\end{eqnarray}
This transformation removes the subtraction terms and suppresses the higher-state contributions.
The resulting LCSR for the product of the strong coupling
and decay constants then reads
\ba
f_H f_{H^*}\,g_{H^*H\pi}
&=& \frac{1}{m_H^2 m_{H^*}}\exp
\left(\frac{m_{H}^2}{M_2^2}+\frac{m_{H^*}^2}{M_1^2}\right)
\nonumber\\
&\times&\iint \limits^{\,\,\,\,\Sigma_0} ds_2 \, ds_1
\, \exp\left(-\frac{s_2}{M_2^2}-\frac{s_1}{M_1^2}\right)
\rho^{\rm (OPE)} (s_1, s_2)\,.
\label{eq:SR2}
\ea
The above sum rule yields the desired $H^*H\pi$ strong coupling,
after dividing out the decay constants of $H^*$ and $H$. For the latter
we will use the two-point QCD sum rules with the same NLO accuracy
and the recent lattice QCD  results.

\section{Double spectral density of the correlation function}

In this section, we  derive the double spectral density
  $\rho^{\rm (OPE)}(s_1,s_2) $ of the correlation function (\ref{eq:corr})
calculated from the light-cone OPE.
We will use the results presented in detail in \cite{DKMMO}.
The procedure to obtain the double spectral density was originally used in \cite{BBKR} at LO,
including the twist-2, 3, 4 contributions.
In \cite{KRWY_2}, the NLO, $O(\alpha_s)$ correction
to the twist-2 contribution was added to the double spectral density.
The result was deduced from the NLO
correction to the twist-2 term of the correlation function obtained
in \cite{KRWY1} (see also \cite{Bagan:1997bp}).
The new element to be included in our calculation is the NLO
correction to the twist-3 part of
$\rho^{\rm (OPE)}(s_1,s_2)$. Apart from that, here we derive the double spectral
density at LO in  a more universal form, valid
for any polynomial structure of the pion DA. We also use the updated
nomenclature of the pion
twist-4 DAs  which differs from the  one  in \cite{BBKR}.
\subsection{Double spectral density at LO}

The OPE near the light-cone $x^2\sim 0$
for the correlation function (\ref{eq:corr})
is valid if both external momenta squared $q^2$ and $(p+q)^2$ are far below
the heavy quark threshold $m_Q^2$.
More specifically, to warrant the power counting in the OPE, it is sufficient that
\begin{eqnarray}
m_Q^2-q^2 \sim m_Q^2-(p+q)^2\sim {\cal O}(m_Q\tau) \,,
\end{eqnarray}
where $\tau\gg\Lambda_{QCD}$ does not scale with $m_Q$.
The heavy quark propagating in the correlation function is then highly virtual.
The initial expression (\ref{eq:corr}) is transformed into
\be
F_\mu(q^2,(p+q)^2)=-im_Q\!\int \! d^4x e^{iqx}\langle \pi(p)| \bar{q}_{1}(x)\gamma_\mu
S_Q(x,0) \gamma_5 q_{2}(0)  |0\rangle\,,
\label{eq:corr1}
\ee
where the heavy quark propagator
$S_Q(x,0)=-i \, \langle 0| T\{Q(x), \, \bar{Q}(0) \} | 0 \rangle $
is expanded near the light-cone.
In the adopted approximation, $S_Q(x,0)$ consists of the
free-quark propagator and  one-gluon emission term.
In the correlation function (\ref{eq:corr1}) with the free heavy-quark
propagator we encounter the vacuum-to-pion
matrix element of the bilocal quark-antiquark operator $\bar{q}_{1}(x)...q_{2}(0)$.
In its turn, the gluon component of the propagator $S_Q(x,0)$ generates the contributions
of the quark-antiquark-gluon operators $\bar{q}_{1}(x)...G_{\mu\nu}(vx)...q_{2}(0)$ with  $0 \leq v \leq 1$.
The emerging vacuum-to-pion matrix elements
are expanded in terms of the pion quark-antiquark (quark-antiquark-gluon)
DAs of growing twist $t=2,3,4$ ($t=3,4$), respectively.
For the leading twist-2 DA we use the well-known standard definition:
\be
\langle \pi^+(p)| \bar{u}(x)\gamma_\mu\gamma_5 d(0) |0\rangle=
-i f_{\pi} \, p_\mu\int\limits_0^1 \!du \,e^{iup\cdot x}\varphi_\pi(u)\,,
\label{eq:tw2}
\ee
where the gauge link has been suppressed for brevity.
All other pion light-cone DAs involved in the
expressions presented below are defined e.g. in \cite{DKMMO}.
With the adopted twist-4 accuracy
the resulting LO expression \cite{DKMMO} for
the invariant amplitude in (\ref{eq:corr1})
represents a sum of the separate twist and multiplicity
contributions:
\begin{align}
F^{\rm (LO)}(q^2,(p+q)^2)=&~\big[F^{\rm (tw2,LO)}
+F^{{\rm (tw3}p,\rm LO)}+F^{\rm (tw3\sigma,LO)}
\nonumber
\\
+&~F^{({\rm tw3},\bar qGq) }+F^{\rm (tw4,\psi)}+F^{\rm (tw4,\phi)}+F^{({\rm tw4},\bar qGq)}\big]
(q^2,(p+q)^2)\,,
\label{eq:OPE}
\end{align}
where the twist-2 contribution is
\begin{align}
F^{\rm (tw2,LO)}(q^2,(p+q)^2)= f_\pi m_Q^2  \int\limits^1_0 \frac{du}{m_Q^2-(q+up)^2}\,\varphi_\pi(u) \,,
\label{eq:OPEtw2}
\end{align}
and the two contributions of the pion two-particle
twist-3 DAs are
\begin{align}
F^{({\rm tw3}p,\rm LO)}(q^2,(p+q)^2)= f_\pi\mu_\pi m_Q \int\limits^1_0 \frac{du}{m_Q^2-(q+up)^2} \,u\,\phi_{3\pi}^p(u)\,,
\label{eq:OPEtw3p}
\end{align}
and
\begin{align}
F^{\rm (tw3\sigma,LO)}(q^2,(p+q)^2)= \frac{f_\pi\mu_\pi}{6}m_Q\int\limits^1_0 \frac{du}{m_Q^2-(q+up)^2}\Bigg(2+\frac{m_Q^2+q^2}{m_Q^2-(q+up)^2}\Bigg) \phi_{3\pi}^\sigma(u)\,.
\label{eq:OPEtw3sig}
\end{align}
The second line in (\ref{eq:OPE}) contains  subleading  contribution of the twist-3 quark-antiquark-gluon DA:
\begin{align}
F^{({\rm tw3},\bar qGq)}(q^2,(p+q)^2)=-4  f_{3\pi} m_Q\int\limits_0^1
\frac{du}{m_Q^2-(q+up)^2}
\Bigg( 1-\frac{m_Q^2-q^2}{m_Q^2-(q+up)^2}\Bigg)\overline\Phi_{3\pi}(u)\,,
\label{eq:OPEtw33}
\end{align}
where we transformed the expression
presented in \cite{DKMMO} into a compact form,
denoting the integrated three-particle DA as $\overline \Phi_{3\pi}(u)$.
The remaining terms in  (\ref{eq:OPE}) contain the twist-4 quark-antiquark  DAs:
\begin{align}
F^{\rm (tw4,\psi)}(q^2,(p+q)^2)= -  f_\pi m_Q^2\int\limits^1_0 \frac{du}{(m_Q^2-(q+up)^2)^2}
\,\,\bar{\psi}_{4\pi}(u)\,,
\label{eq:OPEtw4psi}
\end{align}
\begin{align}
F^{\rm (tw4,\phi)}(q^2,(p+q)^2)= - f_\pi m_Q^4\int\limits^1_0 \frac{du}{2(m_Q^2-(q+up)^2)^3}\,\phi_{4\pi}(u)\,,
\label{eq:OPEtw4phi}
\end{align}
and the integrated linear combinations of the twist-4 quark-antiquark-gluon DAs,
\begin{align}
F^{({\rm tw4},\bar qGq)}(q^2,(p+q)^2)&=  f_\pi m_Q^2\int\limits^1_0
\frac{du}{(m_Q^2-(q+up)^2)^2} \,\overline\Phi_{4\pi}(u)\,.
\label{eq:OPEtw43}
\end{align}

The expressions for all pion DAs and their combinations entering (\ref{eq:OPEtw2})-(\ref{eq:OPEtw43})
are given in Appendix \ref{app:lcda}.
We use the same updated set of twist-3 and 4 DAs from \cite{BBL} as in \cite{DKMMO}.
Their definitions go back to the original work in \cite{Braun:1989iv}.
The form of each DA follows from the conformal  partial-wave  expansion and is given by
a combination of certain orthogonal polynomials in the momentum variables,
such as the variable $u$ in (\ref{eq:tw2}).
The input parameters in  DAs include the overall normalization factors,
e.g., $f_\pi$ in (\ref{eq:tw2}), and the coefficients at the polynomials
normalized at a certain default normalization scale.

Our task is  to derive a double dispersion relation in the form (\ref{eq:ddispOPE})
for each separate term in the OPE (\ref{eq:OPE}).
To this end, we notice  that all contributions of the three-particle DAs
in (\ref{eq:OPE}) have the form of a convolution integral of a single variable $u$,
similar to the contributions of the two-particle DAs.
Moreover, all expressions in (\ref{eq:OPEtw2})-(\ref{eq:OPEtw43}) are
reduced to linear combinations of the two generic integrals
\begin{align}
F^{(\phi)}_{\ell}(q^2,(p+q)^2) \equiv & \int\limits^1_0\! du
\frac{\phi(u)}{\big[m^2_Q-\bar uq^2-u(p+q)^2\big]^\ell}\,,
\nonumber\\
\widetilde F^{(\phi)}_{\ell}(q^2,(p+q)^2) \equiv & \int\limits^1_0 \!du \,
\frac{ q^2\phi(u)\,}{\big[m^2_Q-\bar uq^2-u(p+q)^2\big]^\ell}\,,
\label{eq:genF}
\end{align}
where $\ell=1,2,3$ and $\phi(u)$ has to be replaced by a respective DA, entering (\ref{eq:OPEtw2})-(\ref{eq:OPEtw43}):
$$\phi= \{\varphi_\pi,u\phi^p_{3\pi}, \phi^\sigma_{3\pi},\overline{\Phi}_{3\pi},\bar \psi_{4\pi}, \phi_{4\pi},
\overline{\Phi}_{4\pi}\}.$$
Note that in (\ref{eq:genF}) we have transformed the denominator,
making use of
 $$m_Q^2-(q+up)^2=m^2_Q-\bar uq^2-u(p+q)^2,$$
valid at $p^2=0$ i.e. in a massless pion approximation utilized throughout the paper.
Furthermore, since we aim at the most general
form of the double dispersion relation, it is convenient
to perform a Taylor expansion
of all pion DAs or their integrated combinations
entering (\ref{eq:OPEtw2})-(\ref{eq:OPEtw43})
\begin{align}
\phi(u) = \sum^{\infty}_{k=0} c^{(\phi)}_k \, u^k \,.
\label{eq:expphi}
\end{align}
The expansion (\ref{eq:expphi}) is  convergent for all DAs
including the twist-4 two-particle DA
$\phi_{4\pi}(u)$ in (\ref{eq:OPEtw4phi}), which
contains logarithmic terms of the type $u^k\ln u $ and $\bar{u}^k\ln \bar{u}$ with $k \geq 3$.

Consequently,  it is sufficient to find
the double spectral representation for  the first integral in (\ref{eq:genF})
in which $\phi(u)$ is replaced by the power $u^k$:
\begin{align}
\int\limits^1_0\! du
\frac{u^k}{\big[m^2_Q-\bar uq^2-u(p+q)^2\big]^\ell}
=\int\frac{ds_2}{s_2-(p+q)^2}
\int \frac{ds_1}{s_1-q^2}\,\, \rho_{\ell k}(s_1, s_2)\,,
\label{eq:genFu}
\end{align}
at arbitrary $\ell\geq 1$ and $k>0$, so that the second integral in
(\ref{eq:genF}) is obtained by a simple replacement of
$\rho_{\ell k}(s_1, s_2)$ with $\tilde{\rho}_{\ell k}(s_1, s_2)$ where
\be
\widetilde\rho_{\ell k}(s_1, s_2)=s_1\rho_{\ell k}(s_1, s_2)\,.
\label{eq:rhotild}
\ee
As already said before,  we hereafter neglect the typical subtraction terms
which vanish after the double Borel transformation.
The  formula for the spectral density $\rho_{\ell k}$
can be directly taken from the recent work \cite{Li:2020rcg} where it was derived in a
different context
(see also \cite{Khodjamirian:2011jp} for an alternative technique suitable for
the analogous problems but with the nonvanishing light-hadron mass).
We have:
\begin{align}
\rho_{\ell k}(s_1,s_2) = &
\frac{(-1)^{\ell-1}(-1)^k}{(\ell-1)!\,k!}\,\frac{d^{\,\ell-1}}{d{m^2_Q}^{\ell-1}}\,
\Bigg[\big(s_1-m_Q^2 \big)^k\,
\, \theta\big(s_2-m_Q^2\big)\Bigg]\delta^{(k)}(s_1- s_2)  \,,
\label{eq:rhoelk}
\end{align}
where $\delta^{(k)}(x)\equiv d^k/dx^k[\delta(x)]$.
Note that at $\ell=1$ the expression for $\rho_{1k}(s_1,s_2)$
coincides with the one used in \cite{BBKR}.

The integrals in (\ref{eq:genF}) containing a generic DA $\phi(u)$
can be written as
\begin{align}
F^{(\phi)}_{\ell}(q^2,(p+q)^2)=
\!\int\!\frac{ds_2}{s_2-(p+q)^2}
\int \frac{ds_1}{s_1-q^2}\,\rho^{(\phi)}_{\ell }(s_1, s_2)\,,
\label{eq:rhoell}
\end{align}
and the analogous representation for $\widetilde{F}^{(\phi)}_{\ell}$
with $\widetilde\rho^{(\phi)}_{\ell }(s_1,s_2)$,
where the cumulative spectral densities are obtained combining the
expansion (\ref{eq:expphi})
with the ``elementary'' spectral densities (\ref{eq:rhoelk}),
\begin{align}
\rho^{(\phi)}_{\ell }(s_1, s_2)=\sum^{\infty}_{k=0} c^{(\phi)}_k
\rho_{\ell k}(s_1,s_2),
\qquad
\widetilde\rho^{\,(\phi)}_{\ell }(s_1, s_2)=\sum^{\infty}_{k=0} c^{(\phi)}_k
\widetilde \rho_{\ell k}(s_1,s_2).
\label{eq:exprho}
\end{align}
Replacing  one by one all twist
and multiplicity components in the sum (\ref{eq:OPE})
by their double dispersion forms,
we obtain the double spectral  density for the LO part of the correlation function
\begin{align}
&&\rho^{\rm (LO)}(s_1,s_2)= f_\pi\,m^2_Q \Bigg[
\rho^{(\varphi_\pi)}_1
+ \frac{\mu_\pi}{m_Q} \, \left( \rho^{(u\phi^p_{3\pi})}_1
+\frac{1}{3}\,\rho^{(\phi^\sigma_{3\pi})}_1
+\frac{m^2_Q}{6}\, \rho^{(\phi^\sigma_{3\pi})}_2
+\frac{1}{6} \widetilde\rho^{(\phi^\sigma_{3\pi})}_2 \right)
\nonumber \\
&& + 4 \, \frac{f_{3\pi}}{f_\pi\,m_Q} \,
\left( -\rho^{(\overline{\Phi}_{3\pi})}_1
+m^2_Q\, \rho^{(\overline{\Phi}_{3\pi})}_2
- \widetilde\rho^{(\overline{\Phi}_{3\pi})}_2 \right)
- \rho^{(\bar \psi_{4\pi})}_2
-\frac{m^2_Q}{2} \, \rho^{(\phi_{4\pi})}_3
+\rho^{(\overline{\Phi}_{4\pi})}_2
\Bigg ](s_1,s_2)\,,
\label{eq:rhoLO}
\end{align}
where each term has a form of expansion (\ref{eq:exprho})
with the coefficients $c_k^{(\phi)}$ easily  determined from the polynomial form of the DAs explicitly presented in Appendix \ref{app:lcda}. The expression (\ref{eq:rhoLO}) is new.
Note that it is valid in the chiral limit, i.e. at $p^2=m_\pi^2=0$.
To give useful examples, we present
the contribution to $\rho^{\rm (LO)}$ of the twist-2 and twist-3
DAs
taken in the asymptotic form:
\begin{align}
&\rho_1^{(\varphi_\pi)} (s_1,s_2)= - 6
\left [ (s_1- m_Q^2)\delta^{(1)}(s_1-s_2)+ \frac{1}{2}
(s_1 -m_Q^2)^2\delta^{(2)}(s_1-s_2) \right ] \, \theta(s_2-m_{Q}^2)
\,,
\nonumber\\
&\rho_1^{(u\phi^p_{3\pi})} (s_1,s_2)=-(s_1-m_Q^2)
\delta^{(1)}(s_1-s_2)\, \theta(s_2-m_{Q}^2) \,,
\nonumber \\
& \rho_1^{(\phi^{\sigma}_{3\pi})} (s_1,s_2) = \rho_1^{(\varphi_\pi)} (s_1,s_2) \,,
\nonumber\\
&
\rho_2^{(\phi^{\sigma}_{3\pi})} (s_1,s_2) =
-6  \left [ \delta^{(1)}(s_1-s_2) + (s_1 - m_Q^2)\delta^{(2)}(s_1-s_2) \right ] \theta(s_2 - m_Q^2)
\nonumber \\
& \hspace*{2.5cm} -6 \left [\delta^{(1)}(s_1-s_2) + \frac{1}{2}(s_1 - m_Q^2)\delta^{(2)}(s_1-s_2) \right ] (s_1 - m_Q^2)\delta(s_2-m_Q^2)     \,,
\nonumber\\
&
\tilde{\rho}_2^{(\phi^{\sigma}_{3\pi})} (s_1,s_2) = s_1 \,\rho_2^{(\phi^{\sigma}_{3\pi})} (s_1,s_2)\,.
\label{eq:rhophipi}
\end{align}
The expression (\ref{eq:rhoLO}) enables
to write down the
double spectral representation of $F^{\rm (LO)}$  in a  form (\ref{eq:ddispOPE})
and to perform a double Borel transformation
in a general case of the two unequal parameters $M_1^2,M_2^2$.
In what follows we put $M_1=M_2$ as motivated by the heavy quark symmetry.
After integrating (\ref{eq:rhoLO}) over
the duality region specified in the next subsection,
we will see that the resulting LO part of the sum rule is substantially
simplified and reduced to a linear combination of DAs or their derivatives at the middle point.

\subsection{Double  spectral density at NLO}
To  NLO accuracy, the invariant amplitude we are interested
in the correlation function (\ref{eq:corr}) becomes
\be
F^{\rm (OPE)}(q^2,(p+q)^2)=F^{\rm (LO)}(q^2,(p+q)^2)+
\frac{\alpha_sC_F}{4\pi}F^{\rm (NLO)}(q^2,(p+q)^2)\,,
\ee
where the gluon radiative corrections
at $O(\alpha_s)$ have been calculated in
\cite{KRWY1,Bagan:1997bp}  for the twist-2 part and
in \cite{Ball:2001fp,DKMMO} for the twist-3 part.
The result of this calculation is cast
in a form of the convolution of the hard-scattering amplitudes and the
twist-2 and twist-3 DAs:
\barr
F^{\rm (NLO)}(q^2,(p+q)^2) &=& f_\pi
\int\limits_0^1 du  \Bigg \{ \varphi_\pi (u) \, T_{1}(q^2,(p+q)^2,u)
\nonumber\\
&& + \frac{\mu_\pi}{m_{Q}}\big[
\phi^p_{3\pi}(u) T_{1}^p(q^2,(p+q)^2, u)  +
\phi^\sigma_{3\pi}(u) T_{1}^\sigma(q^2,(p+q)^2,u) \big]\Bigg \},
\hspace{1.2 cm}
\label{eq:convol}
\ea
where the expressions for the twist-2 amplitude
$ T_{1}$, and the twist-3 amplitudes $T_1^p$ and $T_1^{\sigma}$
can be found in \cite{DKMMO}. Here we do not show explicitly
the residual scale dependence of the hard-scattering amplitudes and
of the nonasymptotic parts of pion DAs.
Note that, as explained in \cite{DKMMO},
the twist-3 part of (\ref{eq:convol}) is only
applicable to the asymptotic DAs $\phi^p_{3\pi}(u)$ and $\phi^\sigma_{3\pi}(u)$,
(obtained by putting in (\ref{eq:DAtw3}) the parameter $f_{3\pi}\to 0$)
because the hard-scattering amplitudes $T_1^{p,\sigma}$
are determined perturbatively without taking into account the renormalization-mixing effects
between the two- and three-particle DAs.
Furthermore, in \cite{DKMMO} the NLO part of the correlation function
was represented in a form of
a single-variable dispersion relation, calculating
the imaginary part in $s_2$ which is
the timelike continuation of the variable $(p+q)^2$:
\barr
{\rm Im}_{s_2}F^{\rm (NLO)}(q^2,s_2) &=&
\int\limits_0^1 du \Bigg \{  \varphi_\pi (u) \, {\rm Im}_{s_2} T_{1}(q^2,s_2,u,\mu)
\nonumber\\
&& + \frac{\mu_\pi}{m_{Q}}\bigg[ \phi^p_{3\pi}(u) \,  {\rm Im}_{s_2} T_{1}^p(q^2,s_2,u,\mu)
+ \phi^\sigma_{3\pi}(u) \, {\rm Im}_{s_2} T_{1}^\sigma(q^2,s_2,u,\mu) \bigg] \Bigg \} \,,
\nonumber \\
\label{eq:Ims2}
\ea
at fixed $q^2<m_Q^2$.
The above expression  was used  to derive the NLO terms in LCSRs for the $H\to \pi$ form factors.

Here we need to make a step further and obtain the double spectral density
\be
\rho^{\rm (NLO)} (s_1, s_2) \equiv \frac{1}{\pi^2}
\mbox{Im}_{s_1}\mbox{Im}_{s_2} F^{\rm (NLO)}(s_1,s_2)\,,
\label{eq:rhoopeNLO}
\ee
analytically continuing (\ref{eq:Ims2}) in the variable $q^2\to s_1$.
This double density consists of the three  contributions
stemming from the twist-2 and twist-3 quark-antiquark DAs:
\begin{align}
\rho^{\rm (NLO)}(s_1,s_2)=  \rho^{\rm (tw2,NLO)}(s_1, s_2)+
\rho^{({\rm tw3}p,\rm NLO)}(s_1, s_2)+\rho^{\rm (tw3\sigma,NLO)}(s_1, s_2)\,.
\label{eq:rhoNLO}
\end{align}
We will use the asymptotic DAs for all three NLO terms.
To justify this approximation, we note that at LO the nonasymptotic
contributions due to the Gegenbauer moments in the twist-2 DA
(see (\ref{eq:phipi2}))  contribute at the level of
a few percent to LCSR,  if a typical magnitude of the moments
$a_2,a_4$ is taken (see the section on numerical results below).
An additional $O(\alpha_s)$ factor will suppress these contributions well
below the level of the parametric uncertainties of the sum rule.
For the twist-3 part the nonasymptotic effects at NLO are even
smaller, because already at LO these effects are
determined by a combination of parameters
$f_{3\pi}/(\mu_\pi f_\pi) \sim 0.01$.

For the asymptotic DAs, the calculation of $\rho^{\rm (NLO)}(s_1,s_2)$
simplifies since the integral over $u$ in (\ref{eq:Ims2}) is  performed
before analytically continuing the variable $q^2$ to $q^2=s_1>m_b^2$.
The expressions for the imaginary parts in $(p+q)^2$ of the hard scattering amplitudes
in (\ref{eq:Ims2}) are taken from \cite{DKMMO}.

The twist-2 term in (\ref{eq:rhoNLO}) was already calculated in \cite{KRWY_2}.
We  have recalculated it and confirm the expression presented there.
The resulting expression
for $\rho^{\rm (tw2,NLO)}(s_1, s_2)$ is presented in the Appendix~\ref{app:nlo}.
Note that, since we are now using the $\overline{\rm MS}$ scheme
for the heavy quark mass, an additional $O(\alpha_s)$ piece has to be added
to the expression in \cite{KRWY_2} obtained for the pole mass of the heavy quark.

The derivation of the NLO twist-3 terms in the double spectral density
(\ref{eq:rhoNLO}) is new.
In the course of  calculation we found that the
resulting expressions for $\rho^{({\rm tw3}p,\rm NLO)}(s,s_2)$ and $\rho^{\rm (tw3 \sigma,NLO)}(s_1,s_2)$
contain terms which cancel each other.
Therefore the final expression of the sum of the two
denoted as $\rho^{\rm (tw3,NLO)}(s,s_2)$ is more compact.
It is presented in Appendix~\ref{app:nlo}.

\section{Quark-hadron duality and the sum rule }
\label{sec:QHD}

Having  calculated the double spectral density  (\ref{eq:rhoope}) as
\be
\rho^{\rm (OPE)}(s_1,s_2)= \rho^{\rm (LO)}(s_1,s_2)+ \frac{\alpha_sC_F}{4\pi} \rho^{\rm (NLO)}(s_1,s_2)\,,
\label{eq:rhotot}
\ee
where the LO part is given in (\ref{eq:rhoLO})
and the NLO part represents the sum of the twist-2 and twist-3 parts given,
respectively  in (\ref{eq:tw2nlo}) and (\ref{eq:tw3nlo}),
we are in a position to
perform the integration over a duality region in the LCSR (\ref{eq:SR2}).
In the $\{s_1,s_2\}$ plane, the lower boundary of that region is determined by
the heavy quark threshold (in the chiral limit for light quarks)
and is given by the straight lines $s_1=m_Q^2$ and $s_2=m_Q^2$.
For the  upper boundary symbolized by $\Sigma_0$ in (\ref{eq:SR2})
there is a multiple choice.

As argued in \cite{Neubert:1991sp}, the triangular-type duality region
is preferable in the HQET sum rule for the Isgur-Wise function,
based on the local OPE. This choice was also supported in \cite{Blok:1992fc}
by invoking the double sum rules in nonrelativistic quantum mechanics.
Here we follow the same guidelines in choosing the duality region,
notwithstanding that the LCSR for the $H^*H\pi$ coupling
is based on a different type of OPE, with an interplay of the collinear and soft QCD dynamics.
In \cite{Blok:1992fc} it was shown that duality ansatz works only if the spectral
densities are integrated first over the direction perpendicular
to the diagonal $s_1=s_2$ in the $s_{1, 2}$ plane.
Therefore, we only choose among the regions which process a smooth border crossing
of the diagonal and allow for evaluating the obtained dispersion integrals properly,
implying that the square duality region with a sharp corner on the diagonal
has to be discarded as discussed below.

\begin{figure}[t]
\centering
\includegraphics[width=0.6 \textwidth]{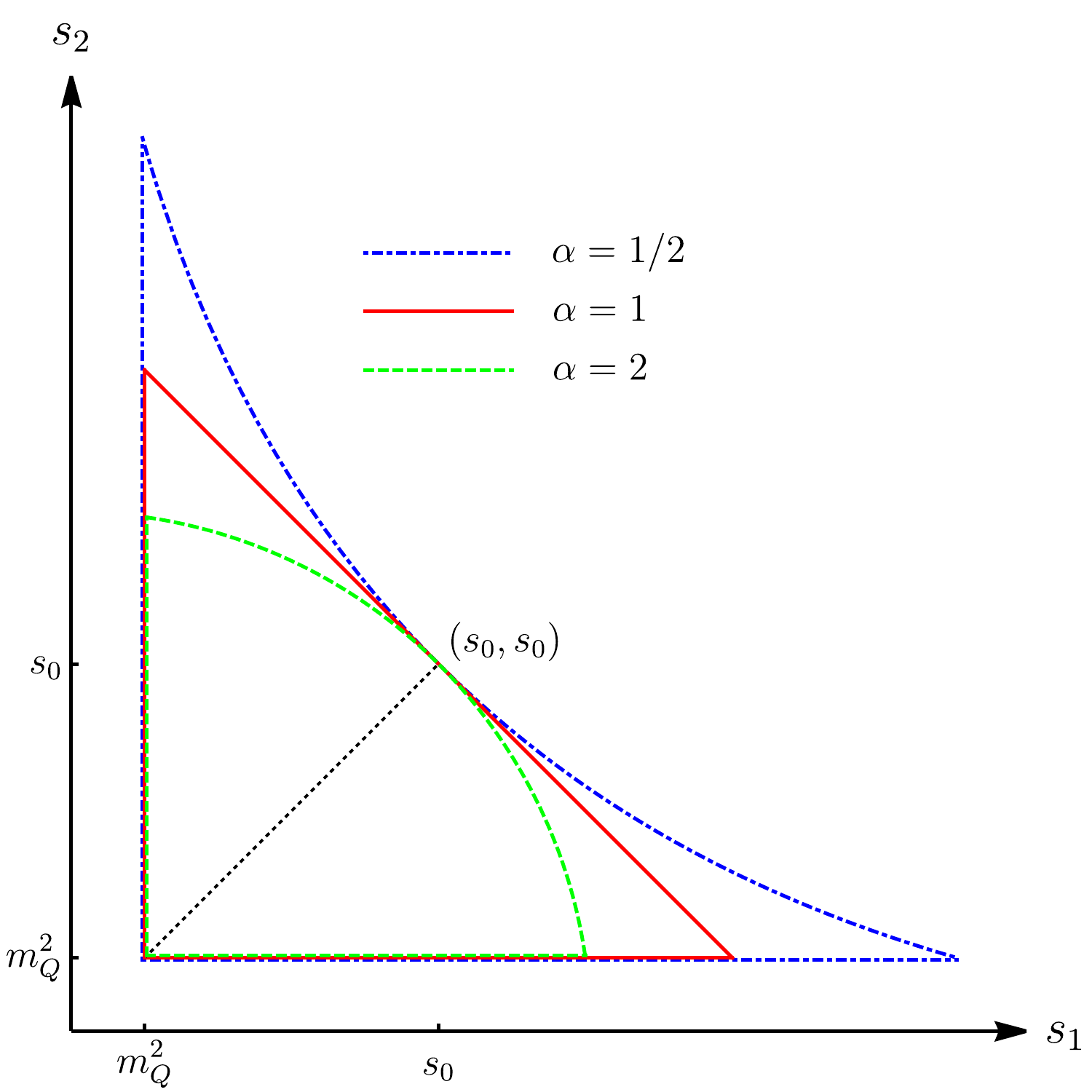}
\caption{The duality regions defined in Eq.(\ref{eq:alpha}).
}
\label{fig:alphas}
\end{figure}

The working duality region includes an interval of the diagonal $s_1=s_2$ 
with a length characterized by the effective threshold $s_0$,
as illustrated  in Figure \ref{fig:alphas}.
The value of this parameter is expected in the ballpark
of the duality threshold in the LCSRs for the $H\to \pi$ form factors.
Our choice for the duality region is motivated by the fact
that the dominant LO part of the spectral density
(\ref{eq:rhotot}) is concentrated near  diagonal,
since $\rho^{\rm (LO)}(s_1,s_2)$
represents a sum of terms proportional to $\delta(s_1-s_2)$
and its first few derivatives.
Due to this property of the LO spectral density, the shape of the
two-dimensional duality region becomes inessential.
However, since we also include
the   $\rho^{\rm (NLO)}(s_1,s_2)$ part, which contains  nonvanishing terms at $s_1\neq s_2$,
a certain dependence on the adopted shape of the duality region will occur.
In order to assess this effect in the NLO part, we will probe the duality regions
with different shapes but possessing the same diagonal interval  along the line $s_1=s_2$.
To this end, it is convenient to use the  parameterization
of the boundaries suggested in \cite{lcsr2}:
\begin{eqnarray}
\left ( {s_1 \over s_{\ast}} \right )^{\alpha} +
\left ( {s_2 \over s_{\ast}} \right )^{\alpha}   \leq 1 \,,
\qquad s_1, \, s_2 \geq  m_Q^2\,.
\label{eq:alpha}
\end{eqnarray}
We will probe the three regions, generated at
\begin{eqnarray}
\label{eq:alph}
\alpha=1\,,  & \qquad &  s_\ast=2s_0\,, \hspace{0.8 cm} {\rm (triangle)};
\nonumber \\
\alpha=\frac{1}{2}\,, & \qquad &   s_\ast=4s_0\,, \hspace{0.8 cm}  {\rm (concave)};
\nonumber \\
\alpha=2\,,  & \qquad &  s_\ast=\sqrt{2}s_0\,,  \hspace{0.5 cm}  {\rm (convex)};
\end{eqnarray}
where $s_\ast$ is adjusted to provide equal diagonal intervals.
These regions are shown, respectively, in Figure~\ref{fig:alphas}.

Note that in the limit $\{\alpha\to \infty,\, s_\ast\to s_0\}$,
the parameterization (\ref{eq:alpha})  represents
a square with the side $s_0$. In this limiting case, the integration of both  NLO twist-2 and twist-3  spectral densities (\ref{eq:tw2nlo}) and (\ref{eq:tw3nlo}) develops a spurious divergence at the vertex $s_1=s_2=s_0$ of the square.
This divergence can be traced back to the presence of the terms involving
\begin{eqnarray}
 {d^3  \over d s_1^3} \,\Big ( \ln |s_1-s_2| \Big) \,.
\nonumber
\end{eqnarray}
To avoid such spurious divergences, it is sufficient to replace
the outmost vertex of the square duality region with a smooth, infinitesimally small curve.
It is clear that the terms in the NLO spectral density containing $\delta(s_1-s_2)$ and its
derivatives, after integration over any of the duality regions defined by (\ref{eq:alpha})  and (\ref{eq:alph}) and shown in Figure~\ref{fig:alphas} yield equal contributions.
Consequently, only those contributions  at the NLO, Eqs. (\ref{eq:tw2nlo}) and (\ref{eq:tw3nlo}), which
do not contain the delta-function  and its derivatives are
 sensitive to the choice of the duality region.

Apart from this, presumably minor effect, which we will numerically study
in the next section, the whole LO and the main part of
NLO  contributions originate from the integration over the
interval on the diagonal which is equal for all duality regions.

Hence, we hereafter adopt the most convenient choice:
the triangular region, satisfying the condition
\begin{align}
\label{eq:tridual}
s_1+s_2 \leq  2 \, s_0.
\end{align}
Returning to the LCSR (\ref{eq:SR2}), we subsequently  assume equal Borel
parameters $$M_1^2=M_2^2=2M^2$$
and rewrite the sum rule as
\begin{align}
f_H f_{H^*}\,g_{H^*H\pi}
= \frac{1}{m_H^2 m_{H^*}}\exp
\left(\frac{m_{H}^2+m_{H^*}^2}{2M^2}\right)
\bigg[\mathcal F^{\rm (LO)}(M^2,s_0) + {\alpha_s \, C_F \over 4 \pi} \,
\mathcal F^{\rm (NLO)}(M^2,s_0)\bigg]\,,
\label{eq:SR3}
\end{align}
introducing the compact notation for the integrals over the triangular duality region,
\begin{align}
\mathcal F^{\rm (LO),(NLO)}(M^2,s_0)\equiv &
\int\limits^{\infty}_{-\infty} \!d s_1
\int\limits^{\infty}_{-\infty} \!d s_2\,
\theta(2s_0-s_1-s_2)\,
\exp\left(-\frac{s_1+s_2}{2\,M^2}\right)
\rho^{\rm (LO),(NLO)}(s_1,s_2) \,.
\label{eq:FLO}
\end{align}
where the lower limits determined by the heavy quark mass are
implicitly given by the theta functions in the expressions of the
spectral densities.

To calculate the LO part in (\ref{eq:SR3}),
we use (\ref{eq:rhoLO}) where the spectral density $\rho^{\rm (LO)}$
is expressed via contributions of the separate DAs.
We then reduce $\mathcal F^{\rm (LO)}(M^2,s_0)$  to a linear combination of the integrals:
\begin{align}
\mathcal F_{\ell}^{(\phi)}(M^2,s_0) \equiv &
\int\limits^{\infty}_{-\infty} d s_1
\int\limits^{\infty}_{-\infty} d s_2\,
\theta(2s_0-s_1-s_2)\,
\exp\left(-\frac{s_1+s_2}{2\,M^2}\right)
\rho^{(\phi)}_{\ell}(s_1,s_2) \,,
\label{eq:Fell}
\end{align}
where $\phi=\varphi_\pi,u\phi^p_{3\pi},\phi^\sigma_{3\pi}$,  etc.
In addition, we define the similar integrals
$\widetilde{\mathcal F}_{\ell}^{(\phi)}(M^2,s_0)$ over $\widetilde \rho^{(\phi)}_{\ell}$.
It is now straightforward to replace each DA by its  Taylor expansion
(\ref{eq:expphi}) and expand the density $\rho^{(\phi)}_{\ell}$ in the elementary components according to (\ref{eq:exprho}).
In fact, in the case of triangular duality region the resulting
formulas for the integrals $\mathcal F_{\ell}^{(\phi)}$ and
$\widetilde{\mathcal F}_{\ell}^{(\phi)}$
can be written in a universal form valid for a generic DA.
To this end, following \cite{BBKR}, we transform
the integration variables in (\ref{eq:Fell}):
\begin{align}
s_1=s(1-v)\,, ~~ s_2=s v
\label{eq:transf}
\end{align}
or, inversely
$$
s = s_1+s_2\,,
\qquad
v = \frac{s_2}{s_1+s_2}\,,
$$
so that  $(s_1-s_2)\to s(1-2v)$, allowing us to integrate out
the $\delta(s_1-s_2)= \delta(1-2v) / s$ functions
and their derivatives over $v$. On the other hand, the exponential factor
in (\ref{eq:Fell}) becomes independent of $v$. As a result, the Taylor expansion
of an arbitrary DA $\phi(u)$ reduces to its value or its derivative at $u=1/2$
and we obtain
\begin{eqnarray}
\label{eq:Fellphi}
&& \mathcal F_{\ell}^{(\phi)}(M^2,s_0)
\nonumber \\
&&  = \frac{(-1)^\ell}{(\ell-1)!}
\bigg\{(-1)^\ell \big(M^2\big)^{2-\ell} \,
\exp \Big(-\frac{m^2_Q}{M^2}\Big)
+ \delta_{\ell 1} \, M^2 \, \exp \Big(-\frac{s_0}{M^2}\Big) \bigg\} \,
\phi(u) \,\bigg|_{u=\frac{1}{2}} \,,
\nonumber \\
&& \widetilde{\mathcal F}_{\ell}^{(\phi)}(M^2,s_0)
\nonumber \\
&& =  -\frac{(-1)^\ell}{2(\ell-1)!}\,\frac{d^{\ell-1}}{d{m^2_Q}^{\ell-1}}
\int_{2\,m^2_Q}^{2\,s_0}ds \,\exp  \left (-\frac{s}{2\,M^2} \right )\,
\left [ u\, \left (\frac{s}{2}-m^2_Q \right ) \, \phi^{\prime}(u)
+ \frac{s}{2} \, \phi(u) \right ]
\bigg|_{u=\frac{1}{2}} \,. \hspace{0.8 cm}
\end{eqnarray}
As we will see below, only  $\widetilde{\mathcal F}_{2}$
contributes, hence for convenience we quote the second integral
in (\ref{eq:Fellphi}) at $\ell=2$
\begin{align}
\widetilde{\mathcal F}_{2}^{(\phi)}(M^2,s_0)=
m^2_Q \,
\exp \Big(-\frac{m^2_Q}{M^2}\Big) \, \phi(u)
+M^2\,\Big[\exp \Big(-\frac{m^2_Q}{M^2}\Big)
- \exp \Big(-\frac{s_0}{M^2}\Big)\Big] u\,\phi'(u)
\bigg|_{u=\frac{1}{2}}\,.
\end{align}
The above formulas are also valid for the twist-4 DA $\phi_{4\pi}$
which contains the specific
$u^k \ln u$ and $\bar u^k \ln \bar u$ terms with $k\geq 3$.

Finally, the LO part of the LCSR  in (\ref{eq:SR3}) is obtained
in a form of a linear combination of the separate DA contributions:
\begin{align}
\mathcal F^{\rm (LO)}(M^2,s_0)=
 f_\pi\,m^2_Q \,\Big[ {\mathcal F}_1^{(\varphi_\pi)}
 + \frac{\mu_\pi}{m_Q} \,{\mathcal F}_1^{(u\phi^p_{3\pi})}
+ \frac{1}{6} \, \frac{\mu_\pi}{m_Q}\,
\Big( 2\,{\mathcal F}_1^{(\phi^\sigma_{3\pi})}
+m^2_Q\, {\mathcal F}_2^{(\phi^\sigma_{3\pi})}
+ {\widetilde{\mathcal F}}_2^{(\phi^\sigma_{3\pi})} \Big)&
\nonumber \\
- 4 \, \frac{f_{3\pi}}{ f_\pi\,m_Q} \,
\Big( {\mathcal F}_1^{(\overline{\Phi}_{3\pi})}
-m^2_Q\, {\mathcal F}_2^{(\overline{\Phi}_{3\pi})}
+ {\widetilde{\mathcal F}}_2^{(\overline{\Phi}_{3\pi})} \Big)
- {\mathcal F}_2^{(\bar \psi_{4\pi})}
-\frac{m^2_Q}{2}\, {\mathcal F}_3^{(\phi_{4\pi})}
+ {\mathcal F}_2^{(\overline{\Phi}_{4\pi})}\Big](M^2,s_0) &\,.
\end{align}
Using (\ref{eq:Fellphi}), we obtain a compact
explicit expression which is straightforward
to use in the numerical analysis of
the LCSR (\ref{eq:SR3}):
\begin{align}
\label{eq:LOtot}
{\mathcal F}^{\rm (LO)}(M^2,s_0)
=& ~ f_\pi \, m^2_Q\,\bigg\{
M^2\,\Big[\exp\Big(-\frac{m^2_Q}{M^2}\Big) - \exp\Big(-\frac{s_0}{M^2}\Big) \Big] \Big[
\varphi_\pi(u)
\nonumber \\
+&~\frac{\mu_\pi}{m_Q} \,\Big(
u\,\phi^p_{3\pi} +\frac{1}{3}\,\phi^\sigma_{3\pi}
+ \frac{1}{6}\,u\,\frac{d\phi^{\sigma}_{3\pi}}{du}\Big)(u)
- \frac{4\,f_{3\pi}}{f_\pi\,m_Q} \,\Big(
\overline{\Phi}_{3\pi}
+ u\,\frac{d\overline{\Phi}_{3\pi}}{du}\Big)(u) \Big]
\nonumber \\
+&~ \exp\Big(-\frac{m^2_Q}{M^2}\Big)\,\Big[
\frac{\mu_\pi\,m_Q}{3} \, \phi^\sigma_{3\pi}
- \bar \psi_{4\pi}
-\frac{1}{4}\,\frac{m^2_Q}{M^2}\, \phi_{4\pi}
+\overline{\Phi}_{4\pi}
\Big](u)
\bigg\}\bigg|_{u=\frac{1}{2}}\,.
\end{align}
Comparing term by term this expression with the one obtained in \cite{BBKR}, we found that
they coincide, although
no explicit duality subtraction was applied to the twist-4 terms
in \cite{BBKR}.
In fact, the peculiar feature of the latter terms is that at equal Borel parameters
the $s_0$-dependent terms vanish, as one can realize using the expressions
for the double spectral density derived here and valid for a generic Taylor-expandable DA.

It remains to obtain the NLO part of (\ref{eq:SR3}).
We have to calculate ${\cal F}^{\rm (NLO)}(M^2, s_0)$ defined
in (\ref{eq:FLO}) by substituting the sum of the twist-2
and twist-3 NLO double spectral densities
$\rho^{\rm (tw2, NLO)}(s_1, s_2)$ and $\rho^{\rm (tw3, NLO)}(s_1, s_2)$
presented in (\ref{eq:tw2nlo}) and  (\ref{eq:tw3nlo}) of  Appendix~\ref{app:nlo}.

The resulting expressions of ${\cal F}^{\rm (NLO)}(M^2, s_0)$
for the triangular duality region reads:
\barr
\mathcal F^{\rm (NLO)}(M^2,s_0) &=& f_{\pi} \, m_Q^2 \,
\int\limits _{2 m_Q^2}^{2 s_0} ds \exp\left(- \frac{s}{2 M^2}\right)
\nonumber \\
&& \times \left  [ f^{\rm (tw2)} \left ( \frac{s}{m_Q^2} -2 \right )
+\left (\frac{\mu_\pi}{m_Q} \right ) f^{\rm (tw3)} \left ( \frac{s}{m_Q^2} -2 \right ) \right ] \,,
\label{eq:NLOBorel}
\ea
with the NLO contributions of twist-2 and twist-3:
\begin{align}
f^{\rm (tw2)}(\sigma) =&~
3\,\Big( 3\,\ln\frac{m^2_Q}{\mu^2} -4 \Big)\,
\Big[\delta(\sigma-0^+) - \frac{1}{2} \Big]
+\frac{\pi^2}{2}
\nonumber \\
&
+6 \,\text{Li}_2\left(-\frac{\sigma }{2}\right)
- 3\,\text{Li}_2\left(-\frac{\sigma }{\sigma +2}\right)
+3\, \text{Li}_2\left(\frac{\sigma }{\sigma +2}\right)
\nonumber \\
&
+\ln\frac{\sigma }{2}\, \Big[
\,3 \,\ln \frac{\sigma+2}{2}
-\frac{3}{2} \,
\frac{\sigma \,(\sigma +4)\,(3 \,\sigma +10)+24}
{(\sigma +2)^3}\Big]
\nonumber \\
&
+ 6\,\ln (\sigma +1)\frac{\sigma \,(\sigma +1)}{(\sigma +2)^3}\,
- \frac{3}{4}\, \frac{3\, \sigma ^2+20 \,\sigma +20}
{ (\sigma +2)^2}
\,,
\label{eq:ftw2}
\end{align}

\begin{align}
f^{\rm (tw3)}(\sigma) =&~
\Big( 3\,\ln\frac{m^2_Q}{\mu^2} -4 \Big)\,
\big[\delta(\sigma-0^+)+2\,\delta'(\sigma-0^+)\big]
+\Big(\frac{4}{3}\,\pi^2 +1\Big)\,\delta(\sigma-0^+)
+\frac{\pi^2}{3}
\nonumber\\
&
+4\, \text{Li}_2\left(-\frac{\sigma }{2}\right)
-2\,\text{Li}_2\left(-\frac{\sigma }{\sigma +2}\right)
+2\, \text{Li}_2\left(\frac{\sigma }{\sigma +2}\right)
\nonumber\\
&
+\ln\frac{\sigma }{2}\,
\Big[ \,2\,\ln\frac{\sigma +2}{2}
+\frac{\sigma ^2+4}{2 \, (\sigma +2)^2}
\Big]
+4 \ln (\sigma +1)
\frac{\sigma ^2+2 \,\sigma +2}{\sigma\,(\sigma +2)^2}
\nonumber\\
&
+\ln \frac{\sigma+2}{2}\left(\frac{\sigma }{8}-\frac{2}{\sigma}\right)
+\frac{3 \,\sigma ^3 +4\,\sigma ^2 -16 \,\sigma -16}
{16 \,(\sigma +2)^2}
\,,\label{eq:ftw3}
\end{align}
where ${\rm Li}_2(x)$ is the  Spence function.
The expression for twist-2 part exactly matches the one given in \cite{KRWY_2}, whereas
the expression of the twist-3 NLO correction (\ref{eq:ftw3})
obtained in the $\overline{\rm MS}$ scheme is a new result.
To switch to the pole-mass scheme for the heavy quark,
it is sufficient to add to (\ref{eq:ftw2}) and (\ref{eq:ftw3})
the terms $\Delta f^{\rm (tw2)}(\sigma)$ and $\Delta f^{\rm (tw3)}(\sigma)$, respectively,
given in (\ref{eq:deltapole}).
As an additional check of our results, we have explicitly verified that
the factorization-scale independence of both the twist-2 and twist-3 terms  in the LCSR
(\ref{eq:SR3}) at $O(\alpha_s^2)$ in the asymptotic limit.

The LCSR (\ref{eq:SR3}) for the strong $H^*H\pi$ coupling,
where $H=B$ or $D$ and, respectively, $m_Q=m_b$ or $m_c$
with the LO and NLO terms given in (\ref{eq:LOtot}) and (\ref{eq:NLOBorel})
is now complete for the triangular duality region and ready for the numerical analysis.

\section{Numerical results}

\begin{table}[h]
\centering
\renewcommand{\arraystretch}{1.5}
\resizebox{\columnwidth}{!}{
\begin{tabular}{|c|c|c|c|}
\hline
parameter & input value & [Ref.]& rescaled values
\\[1mm]\hline
\multicolumn{4}{|c|}{quark-gluon coupling and quark masses}
\\[1mm]\hline
\multirow{2}{*}{$\alpha_s(m_Z)$} &
\multirow{2}{*}{$0.1179 \pm 0.0010$} &
\multirow{6}{*}{\cite{PDG}} &
$\alpha_s(1.5\,\mbox{GeV})=0.3479^{+0.0100}_{-0.0096}$
\\
~& ~& ~&
$\alpha_s(3.0\,\mbox{GeV})=0.2531^{+0.0050}_{-0.0048}$
\\[1mm]
$\overline{m}_c(\overline{m}_c)$    &  1.280 $\pm$ 0.025 \GeV &
~ &
$\overline{m}_c(1.5\,\mbox{GeV})=1.202 \pm 0.023$ \GeV
\\[1mm]
$\overline{m}_b(\overline{m}_b)$   &    4.18  $\pm$ 0.03 \,\GeV &
&$\overline{m}_b(3.0\,\mbox{GeV})=4.46  \pm 0.04$  \GeV
\\[1mm]
\multirow{2}{*}{($\overline{m}_u+\overline{m}_d$)(2 \,\GeV) }
&
\multirow{2}{*}{$6.78 \pm 0.08 $ \MeV}& \multirow{2}{*}{\cite{Aoki:2019cca, PDG}} & ($\overline{m}_u+\overline{m}_d$)(1.5 \,\GeV) =
7.40 $\pm$ 0.09 \MeV\\
~ & ~& ~&
($\overline{m}_u+\overline{m}_d$)(3.0 \,\GeV) =
6.14 $\pm$ 0.07 \MeV\\[1mm] \hline
\multicolumn{4}{|c|}{condensates}
\\[1mm]\hline
 \multirow{2}{*}{$\langle \bar q q \rangle(2\, \GeV)$}
 &
\multirow{2}{*}{$-\left(286\pm 23 \; \MeV \right)^3$}
 &
\multirow{2}{*}{\cite{Aoki:2019cca}}
 &$\langle \bar q q \rangle(1.5\, \GeV)=
 -\left(279\pm 22 \; \MeV \right)^3$
 \\
~& ~& ~&
$\langle \bar q q \rangle(3.0\, \GeV)=
 -\left(295\pm 24 \; \MeV \right)^3$
 \\[1mm]
 $\langle  GG \rangle$ &  $0.012^{+0.006}_{-0.012}\; \GeV ^4$ ~&\multirow{3}{*}{\cite{Ioffe:2002ee}}&-\\[1mm]
$m_0^2$ & 0.8 $\pm$ 0.2 GeV$^2$ & &-\\[1mm]
$r_{vac}$& $0.55\pm 0.45$& &-\\[1mm]
\hline
\end{tabular}}
\caption{QCD parameters used in the LCSRs and two-point sum rules.}
\label{tab:QCD}
\end{table}
To extract the strong couplings $g_{D^*D\pi}$ and $g_{B^*B\pi}$
from the  LCSR (\ref{eq:SR3}),
we need to divide out the decay constants
of the pseudoscalar and vector heavy-light mesons.
Here we will use two different procedures.
The first one, applied in many LCSR applications, prescribes that,
instead  of adopting the fixed  numerical values,
one substitutes in (\ref{eq:SR3})  the two-point QCD sum rules
for decay constants $f_H$ and $f_{H^*}$ ($H=D,B$). These sum rules
presented in Appendix~\ref{app:2pt} are taken from \cite{GKPR}.
For consistency, following the arguments presented in \cite{BBKR,KRWY_2},
the two-point sum rules are taken \footnote{Note that in the previous analysis \cite{KRWY_2}
the perturbative correction to the quark condensate contribution
in the two-point sum rules was absent and is included now. }
 at  NLO, enabling a
partial cancellation of perturbative corrections on both sides of (\ref{eq:SR3}).
As a second, independent option,
we will use the lattice QCD values for the charmed and bottom meson decay constants.
Specifically, we will employ the latest $N_f=2+1+1$ results:
the averages for the heavy pseudoscalar mesons from
\cite{Aoki:2019cca}
and the ratios of the vector and pseudoscalar meson decay constants
obtained in \cite{Lubicz:2017asp},
\begin{eqnarray}
&f_D=212.0\pm 0.7 \,\MeV, \qquad  & f_B=190.0\pm 1.3 \,\MeV ,
\nonumber\\
&f_{D^*}/f_{D}=1.078\pm 0.036,  \qquad & f_{B^*}/f_{B}=0.958\pm 0.022.
\label{eq:lattVP}
\end{eqnarray}

Furthermore, we have to specify the  parameters entering the
LCSR (\ref{eq:SR3}) and the auxiliary two-point sum rules
for decay constants.
The QCD input, including the quark-gluon coupling, the quark masses in $\overline{\rm MS}$ scheme
and the vacuum condensate densities, is listed in Table~\ref{tab:QCD}.
We adopt a very precise value of the light-quark mass combination $(m_u+m_d)$ determined in lattice QCD
\cite{Aoki:2019cca} (see  the average in the quark-mass review of \cite{PDG}).
We also adopt the current interval of the quark condensate \cite{Aoki:2019cca} which is consistent
with the Gell-Mann-Oakes-Renner relation.
The running of the QCD coupling and quark masses is performed with
the four-loop accuracy \cite{Chetyrkin:2000yt} and the matching scales between
$n_f=5 \,(n_f=4)$ and $n_f=4\,(n_f=3)$ are, respectively  4.2 GeV
and 1.3 GeV.

Let us discuss now our choice for the input parameters of pion DAs.
In the LO part (\ref{eq:LOtot}) of the LCSR, we encounter the values of the DAs or
their derivatives at the middle point $u=1/2$. Note that
the midpoint value of a given DA is determined
by a complete set of the coefficients in the conformal expansion, so that  the LCSR
(\ref{eq:SR3}) provides an additional source
of information on the structure of DAs. In this respect it is different from the LCSRs
for the $B\to\pi$ and $D\to \pi$ form factors, where the pion DAs are weighted
by the Borel exponent and integrated over the duality interval.
On the other hand, since the NLO part of the LCSR is calculated for the asymptotic twist-2 and twist-3
two-particle DAs, the only inputs necessary
for a numerical evaluation of (\ref{eq:NLOBorel})
are the normalization factors of these DAs
given, respectively, by the pion decay constant $f_\pi$ and
by the parameter $\mu_\pi$ defined in (\ref{eq:mupi}).

The key parameter of the LO twist-2 part of
LCSR (\ref{eq:SR3}) is the value of $\varphi_\pi(1/2,\mu)$.
Expanding this DA in the Gegenbauer polynomials according to
(\ref{eq:phipi2}), we find
\ba
\varphi_\pi(1/2,\, 1 \,\mbox{GeV})= 1.5
- 2.25\, a_2 + 2.8125 \,a_4  - 3.28125 \,a_6 + 3.69141\, a_8  + \dots\,.
\label{eq:phipi12}
\ea
Hereafter, unless the renormalization scale $\mu$ is explicitly shown,  we denote by $a_n$
the  Gegenbauer moments taken at the scale $\mu=1\,\mbox{GeV}$.
We see that the midpoint value of the twist-2 DA contains
a sign alternating series of all Gegenbauer moments
with slowly growing numerical coefficients.
At larger scales, the moments decrease,  e.g.:
\ba
\varphi_\pi(1/2,\, 3 \,\mbox{GeV})\simeq 1.5
- 1.471\,a_2 + 1.515\, a_4  - 1.553 \,a_6 + 1.585 \,a_8  + \dots\,,
\label{eq:phipi123}
\ea
where the scale dependence calculated using (\ref{eq:rescal})
is absorbed in the numerical coefficients. At $\mu\to \infty$, the value of
$\varphi_\pi(1/2)$ approaches its asymptotic limit equal to 3/2.
Still, at  finite scales, $\varphi_\pi(1/2)$
is an important indicator of the nonasymptotic effects, complementing
the available knowledge of the lowest Gegenbauer moments.

Currently, only  the second moment  $a_2$ of the pion DA is accessible in QCD on the lattice.
We will use the latest quite accurate result:
\be
a_{2}(2 \,\GeV)= 0.116^{+0.019}_{-0.020}
\label{eq:a2latt}
\ee
obtained in \cite{Bali:2019dqc}. From the same analysis,
higher Gegenbauer moments cannot
be extracted reliably, e.g. for $a_4$  only a preliminary
value is quoted,
which will not be considered here.
To estimate and/or constrain the values of $a_{n\geq 4},$
one has to resort to the phenomenologically viable models of $\varphi_\pi(u)$
expanding them in Gegenbauer polynomials.

\begin{table}[t]
\centering
\renewcommand{\arraystretch}{1.5}
\resizebox{\columnwidth}{!}{
\begin{tabular}{|c|c|c|c|c|}
\hline
Twist &Parameter & input value & Source [Ref.] & rescaled values
\\[1mm]\hline
\multirow{5}{*}{2} & $f_\pi$    & $130.4$ MeV    & \cite{PDG} & --
\\[1mm]
~&\multirow{2}{*}{$\varphi_{\pi}(1/2, 2 \,\GeV)$} & \multirow{2}{*}{$1.31\pm 0.03$}  &
\multirow{2}{*}{Model 1~\cite{Bali:2019dqc}}& $\varphi_{\pi}(1/2, 1.5 \,\GeV)= 1.31^{+0.03}_{-0.02}$  \\
~&~&~&~&$\varphi_{\pi}(1/2, 3.0 \,\GeV)= 1.34^{+0.02}_{-0.02}$\\[1mm]
~& \multirow{2}{*}{$\varphi_{\pi}(1/2, 1\,\GeV)$}&
\multirow{2}{*}{$0.99\pm 0.36$}&
\multirow{2}{*}{Model 2~\cite{Cheng:2020vwr}}&
$\varphi_{\pi}(1/2, 1.5\,\GeV)=1.09\pm 0.26$\\
~&~&~&~&$\varphi_{\pi}(1/2, 3.0\,\GeV)=1.18\pm 0.19$\\[1mm]
\hline
\multirow{6}{*}{3} &  \multirow{2}{*}{$\mu_\pi(2\, \GeV)$}&
\multirow{2}{*}{2.87 $\pm$ 0.03 \GeV }& \multirow{2}{*}{$\frac{m_\pi^2}{m_u+m_d}$
\cite{Aoki:2019cca, PDG}} &
$\mu_\pi(1.5\, \GeV)=2.63\,\pm 0.03 \, $ \GeV\\
~&~&~&~&$\mu_\pi(3.0\, \GeV)= 3.17 \pm 0.04$ \GeV
\\[1mm]
~& \multirow{2}{*}{ $f_{3\pi}(1\, \GeV)$} &
\multirow{2}{*}{$(4.5\pm 1.5)\cdot10^{-3}$ GeV$^2$} &  \multirow{4}{*}{\cite{BBL} }&
$f_{3\pi}(1.5\, \GeV)=(3.6\pm 1.2)\cdot10^{-3}$ GeV$^2$
\\
~&~&~&~&$f_{3\pi}(3.0\, \GeV)=(2.8\pm 0.9)\cdot10^{-3}$ GeV$^2$
\\[1mm]
~& \multirow{2}{*}{ $\omega_{3\pi}(1\, \GeV)$} & \multirow{2}{*}{$-1.5 \pm 0.7$} & &
$\omega_{3\pi}(1.5\,\GeV)=-1.2 \pm 0.6$
\\
~&~&~ & ~&$\omega_{3\pi}(3.0\, \GeV)=-1.0 \pm 0.5$\\[1mm]
\hline
\multirow{4}{*}{4} &\multirow{2}{*}{$\delta_\pi^2(1\, \GeV)$}&
\multirow{2}{*}{$0.18 \pm 0.06$ GeV$^2$ } &
\multirow{4}{*}{\cite{BBL}}&
$\delta_\pi^2(1.5\, \GeV) =0.16 \pm 0.05$\\
~&~&~&~&$\delta_\pi^2(3.0\, \GeV) =0.14 \pm 0.05$\\[1mm]
~& \multirow{2}{*}{$\epsilon_{\pi}(1\, \GeV)$ }&
\multirow{2}{*}{$0.5\pm 0.3$} & ~ &
$\epsilon_{\pi}(1.5\, \GeV)=0.4\pm 0.2$\\
~&~&~&~&$\epsilon_{\pi}(3.0\, \GeV)=0.3\pm 0.2$\\
\hline
\end{tabular}
}
\caption{ Parameters of pion DAs.}
\label{tab:piDA}
\end{table}

To choose the input value of $\varphi_\pi(1/2)$, we adopt two such
models. The first one denoted here as Model 1 was suggested in \cite{Bali:2019dqc}:
\be
{\rm Model\; 1:} \qquad \varphi_\pi(u)=\frac{\Gamma(2+2\alpha_\pi)}
{[\Gamma(1+\alpha_\pi)]^2}\,
u^{\alpha_\pi}\,(1-u)^{\alpha_\pi}\,.
\label{eq:model1}
\ee

Its single free parameter  is fixed
by equating the second Gegenbauer moment of this model
to the lattice QCD result (\ref{eq:a2latt}), yielding
$\alpha_\pi(2\,\GeV)=0.585^{+0.061}_{-0.055}$.
In addition, the first inverse moment of this DA is
\be
\int\limits_0^1 \! du \,\frac{\varphi_\pi(u,2\,\GeV)}{1-u}=2+
\frac{1}{\alpha_\pi(2\,\GeV)}=
3.71^{+0.18}_{-0.16}\,.
\ee

The corresponding
 midpoint value of the DA (\ref{eq:model1}) is given in Table~\ref{tab:piDA}.
Note that the inverse moment serves as the main input in the QCD calculation of the
photon-pion transition form factor  \cite{Khodjamirian:1997tk,Agaev:2012tm,Li:2013xna,Wang:2017ijn}.
As noted in \cite{Bali:2019dqc}, applying this method
with the above value, one achieves a good description of data on this form factor.

Our Model 2 is of a different origin and is based on the comparison
of the LCSR for the pion electromagnetic form factor
\cite{Braun:1999uj} with the experimental data. We use the results
of the recent analysis \cite{Cheng:2020vwr}, where a dispersion relation
and the data in the timelike region are used to reproduce the  pion form factor in the spacelike region.
These results are then used to fit the LCSR form factor calculated to the twist-2 NLO accuracy including
the subleading twist-4,6 terms. Among various versions of the fitted twist-2 DAs we choose
the optimal one with the first three moments in the Gegenbauer expansion
(\ref{eq:phipi2}). The fit results obtained in \cite{Cheng:2020vwr} at the scale of 1 GeV are
\be
{\rm Model\; 2:} \qquad a_2=0.270\pm 0.047, ~~a_4=0.179\pm 0.060,~~ a_6=0.123 \pm 0.086\,,
\label{eq:mod2}
\ee
with the correlation matrix:
\be
\Bigg(\begin{array}{ccc}
    1.0   &    -0.15  &   -0.13\\
   -0.15  &1.0 &       -0.13\\
   -0.13  &  -0.13  &  1.0
\end{array}\Bigg)\,.
\label{eq:corrm}
\ee
The corresponding input value of $\varphi_\pi(1/2,  1\,\rm{GeV})$
 is given in Table~\ref{tab:piDA}.

In the same Table we specify the input parameters entering the
 pion twist-3 and twist-4 DAs presented in Appendix~\ref{app:lcda}. These DAs were
 worked out in \cite{BBL} to the next-to-leading order of conformal expansion.
 Their normalization and nonasymptotic coefficients at $\mu=1$ GeV used as an input here were calculated from the two-point QCD sum rules (see \cite{BBL} and references therein).
We notice, in particular, the relative smallness of the twist-3 parameter $f_{3\pi}$,
which determines the nonasymptotic part of the two-particle
DAs and the normalization of the three-particle DA.
Hence, the large twist-3 contribution
to LCSR is, to a good precision, determined by
the asymptotic two-particle DAs $\phi^{p}_{3\pi}$ and $\phi^{\sigma}_{3\pi}$ at the  midpoint.
Here we greatly benefit from the very accurate value of the twist-3 normalization parameter $\mu_\pi$
which is determined by the light-quark masses. Note at the same time that the $O((m_u+m_d)^2/m_\pi^2)$ correction
to the ratio of normalization factors for $\phi^{\sigma}_{3\pi}$ and $\phi^{p}_{3\pi}$
is still small enough to  be neglected safely.
The contributions of the twist-4 two- and three-particle DAs, as we will see, are altogether strongly suppressed.
Therefore, there is no compelling reason to go beyond the current accuracy,
and e.g., calculate the NLO corrections to the twist-4 part, which is technically a challenging task.

To complete the choice of the input, we take the meson masses from \cite{PDG},
considering, for definiteness, the strong coupling
$\langle D^{*+}\pi^-|D^0\rangle$ and, correspondingly, $\langle \bar{B}^{*0}\pi^-|B^-\rangle$
in (\ref{eq:strong}). All other couplings with different combinations of charges are
related to the above ones via the isospin symmetry (see e.g. \cite{BBKR}). Finally, we
specify the variable parameters of the LCSR (\ref{eq:SR3}),
which include: the renormalization scale of the quark-gluon coupling and quark masses,
the factorization scale,  the Borel parameters and the quark-hadron duality thresholds.
Since we perform the calculations at finite masses,
these scales and thresholds are evidently different in the sum rules involving
charmed and bottom mesons. On the other hand,
heavy-quark spin symmetry allows us to equate certain scales,
most importantly, the Borel parameters in the $H$ and $H^*$ channels.
The chosen default values and intervals of all relevant scales and thresholds
are presented in Table~\ref{tab:scales}.
\begin{table}[h]
\centering
\begin{tabular}{|c|c|c|}
\hline
Parameter & default value (interval) & [Ref.]  \\[1mm]
\hline
\multicolumn{3}{|c|}{ charmed meson sum rules}\\
\hline
$\mu$ (GeV) & 1.5~(1.0\,-\,3.0)&
\multirow{3}{*}{\cite{KKMO}}
\\[1mm]
$M^2$ (GeV$^2$)  & 4.5~(3.5\,-\,5.5) &
\\[1mm]
$ s_0$  (GeV$^2$) & 7.0~(6.5\,-\,7.5) &
\\[1mm]\hline
$\bar{M}^2$(GeV$^2$) & 2.0~(1.5\,-\,2.5)&
\multirow{3}{*}{\cite{GKPR}}
\\[1mm]
$\bar{s}_{0}$ (GeV$^2$) & 5.6 &
\\
$\bar{s}_{0}^{\,*}$(GeV$^2$) & {6.2} & \\
\hline
\multicolumn{3}{|c|}{bottom meson sum rules}\\
\hline
$\mu$  (GeV) & {3.0~(2.5\,-\,4.5)}&
\multirow{3}{*}{\cite{KMOW}}  \\
  $M^2$  (GeV$^2$)  & {16.0~(12.0\,-\,20.0)} &\\
$ s_0$  (GeV$^2$) &  {37.5\,(35.0\,-\,40.0)} &  \\
\hline
$\bar{M}^2$ (GeV$^2$)& {5.5~(4.5\,-\,6.5)} &
\multirow{3}{*}{\cite{GKPR}} \\
$\bar{s}_{0}$(GeV$^2$)  & 33.9 & \\
$\bar{s}_{0}^{\,*}(\GeV^2$) & 34.1&  \\
\hline
\end{tabular}
\caption{The renormalization scale $\mu$, Borel parameters $M^2$ and $\bar{M}^2$
and duality thresholds $s_0$ and $\bar{s}_0$ ($\bar{s_0}^{*}$)
used, respectively in the LCSR and the two-point sum rules for the $H$ ($H^*$) decay constants
for both charmed and bottom mesons.}
\label{tab:scales}
\end{table}
Here we follow the numerical analysis of the related LCSRs
for the $D\to \pi$ and $B\to \pi$ form factors. More specifically,
we employ for the charm and bottom cases of (\ref{eq:SR3}) the same
variable scales and thresholds as, respectively, in  \cite{KKMO} and \cite{KMOW}.
The compelling argument is that  we deal here  with the same underlying correlation
function and the same light-cone OPE as in the form factor sum rules.
Also, the renormalization scales of $\alpha_s$ and quark masses are taken equal to
the factorization scale $\mu$ appearing in the OPE of the correlation function
\footnote{Note that the currents $j_{\mu}$ and $j_5$ in (\ref{eq:corr}) are
renormalization invariant.}  (\ref{eq:corr}).
In the adopted approximation the factorization scale
reveals itself  in the nonasymptotic components of DAs in the LO part, while in the NLO part we use the asymptotic  DAs.
For consistency, the same scale is used in the corresponding two-point sum rules for $f_H$ and $f_{H^*}$.

In Tables~\ref{tab:QCD}
and \ref{tab:piDA}, apart from the input values of the scale-dependent
parameters at a given reference scale
$\mu=2.0$ GeV  or $\mu=1.0$ GeV,
we also present, for convenience, their rescaled values at  $\mu=1.5$ GeV  and
$\mu=3.0$ GeV, to be used in the sum rules
with charmed and bottom mesons, respectively.
Note that the midpoint value of the twist-2 DA can be rescaled only if it is
expressed as a linear combination of multiplicatively renormalizable Gegenbauer moments,
because the latter possess different anomalous dimensions.
Hence, for the Model 1 we first calculate the Gegenbauer moments of the DA in (\ref{eq:model1}),
and then, forming the expansion, rescale each moment according to
(\ref{eq:rescal}). For the Model 2 the rescaling is straightforward.
The resulting values of $\varphi_\pi(1/2)$ for both models
are presented in Table \ref{tab:piDA} at the two default scales,
together with the  parameters  used for the twist-3 and twist-4 DAs.
The formulas determining the scale dependence of the latter can be found e.g. in the Appendix A of \cite{DKMMO}.
As already mentioned above, the intervals of Borel parameter, as well as the values of threshold parameters in the LCSR (\ref{eq:SR3})
are the same as in the LCSR analyses  for $D\to\pi$ \cite{KKMO} and $B\to \pi$ \cite{KMOW} form factors. In each of these analyses, the duality threshold was adjusted by
fitting the differentiated sum rule to the heavy meson mass.
The Borel parameters and duality thresholds
in the two-point sum rules for the $H^{(*)}$ decay constants are
taken from \cite{GKPR}.

With the input specified above, we calculate first the product
of the strong coupling and decay constants from the
sum rule (\ref{eq:SR3}). The results are presented
in Table \ref{tab:products}
at the central input and at default scales, including also the separate twist and NLO contributions. The twist-2 and twist-3 contributions are
at the same level, similar as in the LCSRs for heavy-to-light form factors.  In the twist-2 LO part of LCSR (\ref{eq:SR3})
the contributions of nonasymptotic terms
are quite noticeable, as can be seen from
comparison of the results for the two different DA models.
At the same time,
the share of asymptotic DAs
constitutes 93\% (87\%) of the twist-3 LO contribution for the
bottom (charmed) meson case.
The convergence of the OPE is supported by the smallness of
the twist-4 contributions.
In addition, as already mentioned, the twist-5 and twist-6 terms in the
OPE of the correlation function (\ref{eq:corr})
obtained in the factorizable approximation in \cite{Rusov:2017chr}
have negligible impact, allowing us to neglect them here.
On the other hand, as seen from Table \ref{tab:products},
the NLO contributions are appreciable, reaching e.g. for the bottom meson case the level of 20\% (35\%) for twist 2 (twist 3).
The results presented in this Table correspond to our default choice of the triangle
duality region in the $(s_1,s_2)$ plane, described by the parameterization (\ref{eq:alpha}) at $\alpha=1/2$.
In addition, to investigate how the choice of the duality region influences the LCSR,
we calculated the NLO terms at the same default value of $s_0$ for two other choices
of the region corresponding to $\alpha=1$ and $\alpha=2$. The results are presented in Table~\ref{tab:region}. We see that deviations with respect to the choice of
triangle region are at the level of a few percent of the total values
of both $f_{D}f_{D^*}g_{D^*D\pi}$ and $f_{B}f_{B^*}g_{B^*B\pi}$.
We include this deviation into the total uncertainty for the
predicted strong couplings $g_{D^*D\pi}$ and $g_{B*B\pi}$.
\begin{table}[t]
\centering
\setlength\tabcolsep{5pt}
\def\arraystretch{1.2}
\begin{tabular}{|c||c|c|c|c|c||c|}
\hline
LCSR result  & tw~2 LO  & tw~2 NLO & tw~3 LO & tw~3 NLO
&tw~4 & total\\
\hline
$f_D\,f_{D^*}\,g_{D^*D\pi}$
& 0.188 (Model 1)& \multirow{2}{*}{0.049}
& \multirow{2}{*}{0.333} & \multirow{2}{*}{0.115}
& \multirow{2}{*}{-0.001} & 0.684
\\
$[{\rm GeV}^2]$ & 0.156 (Model 2) &~ &~ &~ &~ & 0.652
\\
\hline
$f_B\,f_{B^*}\,g_{B^*B\pi}$
& 0.416  (Model 1) & \multirow{2}{*}{0.081}
&\multirow{2}{*}{0.395} &\multirow{2}{*}{0.148}
&\multirow{2}{*}{-0.004} & 1.037
\\
$[{\rm GeV}^2]$ &0.367 (Model 2) &~ &~ &~ &~ & 0.988
\\
\hline
\end{tabular}
\caption{Numerical results at the central input.}
\label{tab:products}
\end{table}
\begin{figure}[h]
\centering
\includegraphics[width=1. \textwidth]{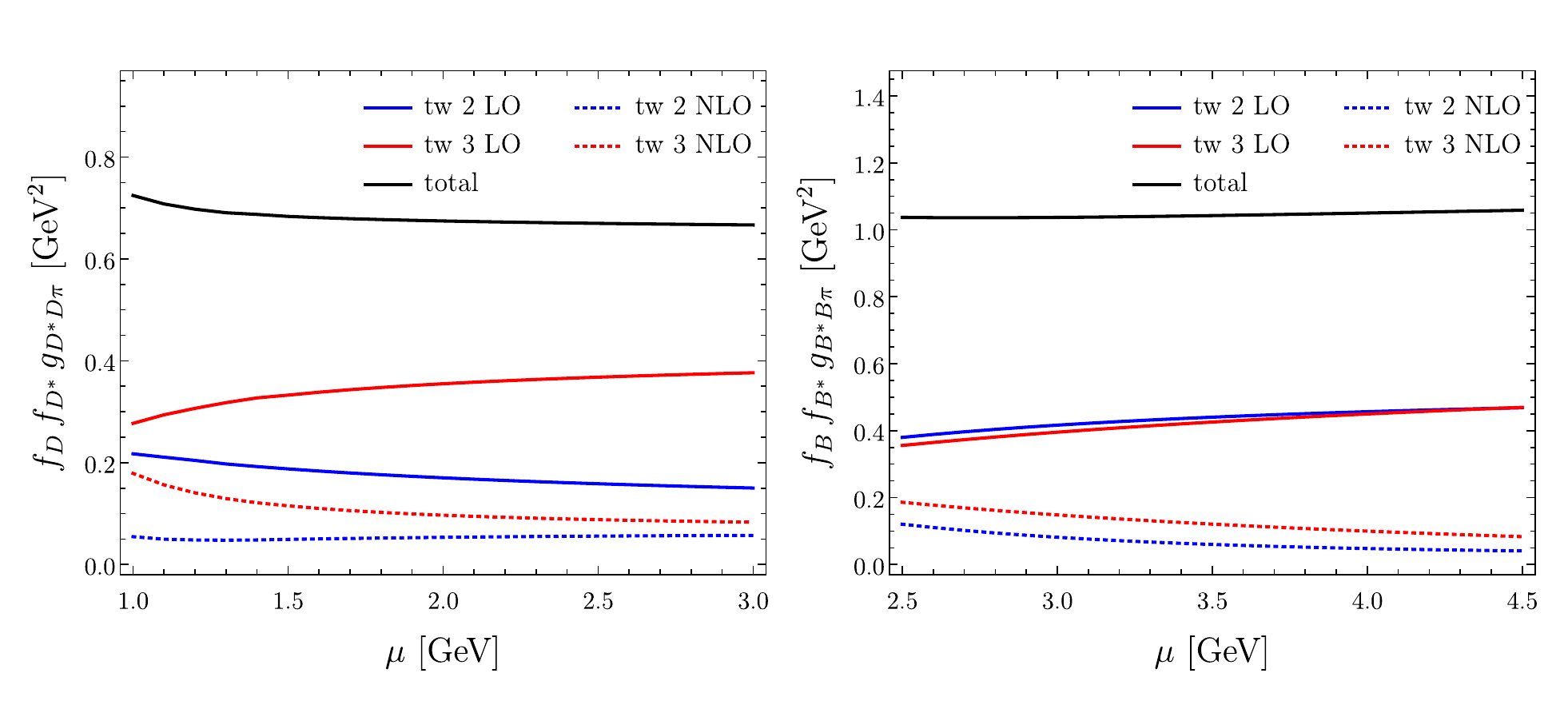}
\caption{Scale dependence of the products
$f_D f_{D^*}g_{D^*D\pi}$ and $f_B f_{B^*} g_{B^*D\pi}$ calculated from LCSR.
Displayed are the total values and separate twist-2 and twist-3 contributions
for Model 1 of the twist-2 pion DA and at central values of all other
input parameters.}
\label{fig:mu}
\end{figure}

\begin{table}[h]
\centering
\setlength\tabcolsep{5pt}
\def\arraystretch{1.1}
\begin{tabular}{|c||c||c|c|}
\hline
LCSR result  & $\alpha$ & tw~2 NLO & tw~3 NLO
\\
\hline
\multirow{3}{*}{$f_D\,f_{D^*}\,g_{D^*D\pi}~[{\rm GeV}^2]$}
& 1/2 &
0.056& 0.128\\
~ & 1 & 0.049 & 0.115 \\
~ & 2 & 0.038 & 0.093\\
\hline
\multirow{3}{*}{$f_B\,f_{B^*}\,g_{B^*B\pi}~[{\rm GeV}^2]$}
& 1/2 & 0.090& 0.158\\
~ & 1 & 0.081 & 0.148 \\
~ & 2 & 0.066 & 0.131 \\
\hline
\end{tabular}
\caption{Numerical results for different duality regions at central values of parameters.}
\label{tab:region}
\end{table}

To finally obtain these couplings, we divide out
the heavy-meson decay constants applying the two different methods described above:
we  either use the two-point sum rules or the lattice QCD results listed in  (\ref{eq:lattVP}).
For comparison, we quote the values of decay constants calculated at NLO from the two-point sum rules at central input:
\be
f_D=190.7~\MeV, \qquad  f_{D^*}= 247.3~\MeV, \qquad
f_B=201.0~\MeV, \qquad  f_{B^*}=214.0~\MeV\,.
\nonumber
\ee

Our final results are shown in Table \ref{tab:res} for both Model 1 and 2 of the twist-2 DA and for both choices of the decay constants.
The interval attributed to each separate entry in Table~\ref{tab:res}
is evaluated, adding in quadrature the separate uncertainties
caused by individual variations of all input parameters and scales
within their adopted intervals in LCSR. When using the
two-point sum rules for the $D^{(*)}$ and $B^{(*)}$ decay constants, we vary
the scale $\mu$ and Borel parameters (condensate densities) concertedly
(independently). The lattice results for the decay constants
have very small errors which play almost no role in the uncertainty budget.
Finally, the total uncertainty quoted in Table~\ref{tab:res}  also
includes the variation due to the change of the duality
region in the NLO part as described in Sec. \ref{sec:QHD}.
We assume that the
latter uncertainty at least partially assesses  the ``systematic error"
of LCSR caused by the quark-hadron duality ansatz.

\begin{table}[h]
\centering
\setlength\tabcolsep{5pt}
\def\arraystretch{1.4}
\begin{tabular}{|c||c||c|c|c|c|}
\hline
$\varphi_\pi(1/2)$ & decay constants & $g_{D^*D\pi}$ & $g_{B^*B\pi}$ & $\hat{g}$ &$\delta$\,[GeV]\\
\hline
\multirow{2}{*}{Model 1} & 2-point sum rule
& $14.5^{+3.5}_{-2.4}$ & $24.1^{+4.5}_{-3.8}$
& $0.18^{+0.02}_{-0.03}$
& $3.28^{+0.62}_{-0.17}$
\\
& Lattice QCD & $14.1^{+1.3}_{-1.2}$
& $30.0^{+2.6}_{-2.4}$
& $0.30^{+0.02}_{-0.02}$
& $1.17^{+0.04}_{-0.04}$\\
\hline
\multirow{2}{*}{Model 2} & 2-point  sum rule
& $13.8^{+3.1}_{-2.3}$ & $23.0^{+4.5}_{-3.8}$
& $0.17^{+0.03}_{-0.03}$
& $3.31^{+0.30}_{-0.01}$\\
& Lattice QCD & $13.5^{+1.4}_{-1.4}$
& $28.6^{+3.0}_{-2.8}$
& $0.29^{+0.03}_{-0.03}$
& $1.18^{+0.00}_{-0.02}$\\
       \hline
\end{tabular}
\caption{LCSR  results for the strong couplings of the charmed and bottom mesons for the two
methods of dividing out the decay constants and the two models of the pion twist-2 DA,
at the central values of parameters.}
\label{tab:res}
\end{table}

The LCSR prediction for $D^*D\pi$ ($B^*B\pi$)  strong coupling has altogether
an estimated uncertainty of 20-25\% (15-20\%),
if the heavy-meson decay constants are replaced by the two-point sum rules.
The uncertainties become smaller when we use  more accurate lattice QCD values
for the heavy-meson decay constants.
For the two options for  decay constants the
predicted intervals
 of  $g_{D^*D\pi}$ are close to each other,
whereas the intervals of $g_{B^*B\pi}$ only marginally agree.
The shift between the central values
of the $B^*B\pi$ coupling  mainly originates due to the ${\cal O} \, (20 \%)$ difference
between the lattice-QCD value of $f_{B^{\ast}}$  and the central value of
the two-point sum rule prediction \footnote{As already mentioned, for consistency,
here we use the two-point sum rules  at NLO.
More accurate sum rules at NNLO yield \cite{GKPR}  the ratio
$f_{B^{\ast}}/f_B=1.02^{+0.02}_{-0.09}$, reflecting the heavy-spin symmetry breaking effect. Within uncertainties,
this result is in agreement with (\ref{eq:lattVP}) calculated in the lattice QCD \cite{Lubicz:2017asp}.}.
We also notice that the choice of the model for the pion twist-2
DA is less important for the charmed-meson strong coupling, the reason being
a dominance of the twist-3 contribution enhanced by the ratio $\mu_\pi/m_c$
with respect to $\mu_\pi/m_b$ for the bottom-meson coupling.

Comparing our numerical results with the original LCSR calculation in \cite{BBKR},
we notice a substantial increase  of  the products of the strong couplings 
and decay constants displayed in Table~\ref{tab:products}
with respect to the results  $f_Df_{D^*}g_{D^*D\pi}=0.51 \pm 0.05~\GeV^2$
and $f_Bf_{B^*}g_{B^*B\pi}= 0.64\pm 0.06~\GeV^2$ obtained in \cite{BBKR}. 
This increase is mainly caused  by the twist-2 and twist-3 NLO terms in the LCSR,
which were absent in \cite{BBKR}.
In addition, the updated input parameters in our numerical analysis differ
from the ones adopted in \cite{BBKR}.
Most importantly, we use the $\overline{\rm MS}$ heavy-quark masses 
instead of the pole-mass scheme employed in \cite{BBKR}. 
In particular, the updated value of the twist-3 normalization parameter $\mu_\pi$ has become substantially larger.
Moreover, the updated inputs affect the numerical LO result in different directions, 
largely compensating each other in the case of charmed mesons
and generating an additional increase in the case of bottom mesons.
The determined strong couplings are numerically influenced by the magnitude 
of the products of heavy-meson decay constants.
In \cite{BBKR} they were taken from the two-point sum rules at LO. 
By contrast, the larger values of these products are employed here 
such that  the above-mentioned increase of the LCSR results 
is either partially (for the charmed-meson case) or almost completely (for the bottom-meson case) compensated, as can be seen by comparing
our predictions presented in Table~\ref{tab:res} with
$g_{D^*D\pi}= 12.5\pm 1.0 $ and $g_{B^*B\pi}= 29\pm 3 $ from \cite{BBKR}.

It is also instructive to investigate the limit of infinitely heavy-quark mass
obtained from the LCSR (\ref{eq:SR3}) for the strong coupling.
This sum rule, where the
underlying correlation function is calculated at the finite mass $m_Q$,
not only  reproduces the leading-power behaviour of the coupling at $m_Q\to \infty$
 but also enables a quantitative  assessment of  the $1/m_Q$ corrections.
For the heavy-to-light form factors obtained from the LCSRs, the heavy-quark mass  expansion has been
investigated in the early papers \cite{Ali:1993vd,Khodjamirian:1998vk}.
To proceed, we apply to the sum rule (\ref{eq:SR3}) the
usual scaling relations (valid up to the inverse heavy-quark mass corrections):
\begin{eqnarray}
f_H=f_{H*}=\frac{\hat{f}}{\sqrt{m_Q}},~~m_H=m_{H*}=m_Q+\overline{\Lambda},~~
M^2= 2m_Q\tau,~~ s_0=m_Q^2+2m_Q\omega_0 \,,
\end{eqnarray}
where $\hat{f}$ and $\overline{\Lambda}$ are, respectively, the static decay constant
and the binding energy of heavy meson in HQET,
and the parameters $\tau$ and $\omega_0$
do not scale with $m_Q$. We obtain at LO
\begin{eqnarray}
g_{H^*H\pi} = \frac{2m_Q }{f_\pi} \Bigg[\frac{f_\pi^2}{\hat{f}^2}e^{\overline{\Lambda}/\tau}\bigg\{\tau\left(1-
e^{-\omega_0/\tau}\right)\varphi_\pi(1/2)+
\frac{\mu_\pi}{6}\phi_{3\pi}^{\sigma}(1/2)
-\frac{\phi_{4\pi}(1/2)}{16\tau}\bigg\}\Bigg] + \dots \,,
\hspace{0.8 cm}
\label{eq:HQET}
\end{eqnarray}
where ellipsis indicates the inverse heavy-mass corrections.
Comparing the above formula with the definition  (\ref{eq:heavy}) and taking the
limit $m_Q\to \infty$, we notice that the expression in square brackets
is nothing but a sum rule for the static coupling $\hat{g}$ in HM$\chi$PT.
Note that the twist-3 and 4 terms also contribute to this sum rule.
Moreover, from the LCSR (\ref{eq:SR3}) we are in a position
to estimate both the static coupling and the inverse mass corrections
to it. Including the NLO terms in this limiting procedure is however nontrivial,
because one has to resum the logarithms of the heavy-quark mass.
A systematic way is to derive the LCSR for the strong coupling directly in HQET, a task which is beyond our scope here.

Expanding the rescaled sum rule (\ref{eq:SR3})
in the powers of $1/m_Q$, we  follow \cite{BBKR},
and parameterize the LCSR result for the strong coupling  a form (\ref{eq:heavy}) with an added inverse heavy-mass correction
\footnote{This parametrization is further supported by the heavy quark expansion
of  the strong couplings $H^{\ast} H \pi$ in the framework of HM$\chi$PT \cite{Boyd:1994pa,Casalbuoni:1996pg}.}:
\be
g_{H^*H\pi}=\frac{2m_H\hat{g}}{f_\pi}\left(1+\frac{\delta}{m_H}\right), ~~~(H=D,B) \,.
\label{eq:hqe}
\ee

Equating the above formula to the obtained values of $g_{D^* D \pi}$ and $g_{B^* B \pi}$ presented in  Table~\ref{tab:res}, we
encounter the two equations yielding the parameters $\hat{g}$ and $\delta$.
 Their resulting values  are presented in the last two columns
of the same table. We find that the inverse mass corrections are large,
especially in the case of the $D^*D\pi$ strong coupling as it was already noticed in \cite{BBKR}.
Hence, estimating the
static coupling $\hat{g}$  from the known
$D^*D\pi$ or $B^*B\pi$ couplings via the relation (\ref{eq:heavy}),
as it is frequently done in the literature, is actually not reliable in practice.

The obtained values for the $D^*D\pi$ coupling   can be compared
to its experimentally measured value, extracted from
the width of the $D^{\ast+}\to D^0 \pi^+$ decay:
\be
\Gamma(D^{*+}\to D^0\pi^+)=\frac{g_{D^*D\pi}^2}{24\pi m_{D^*}^2}|\vec{p}|^3\,,
\label{eq:width}
\ee
where $|\vec{p}|=39 $ MeV is the decay momentum in the rest frame of $D^*$.
The above formula for the partial width is obtained from the
decay amplitude which is defined by crossing-transforming
the initial definition (\ref{eq:strong})
\be
\langle D^0(q-p) \pi^+(p)|D^{*+}(q)\rangle =-g_{D^* D\pi}\,p^\mu\epsilon_\mu^{(D^*)}\,.
\label{eq:strongD}
\ee
The PDG average \cite{PDG} of the two measurements \cite{CLEODstDpi,BaBarDstDpi}
of the $D^{*\pm}$-meson total width is
$\Gamma_{\rm tot}(D^{*\pm})= 83.4 \pm 1.8$ keV
\footnote{For the $D^{*0}$ total width only an upper limit is measured so far. To relate the total widths
of charged and neutral $D^*$ mesons, the isospin symmetry
is not sufficient, because one needs in addition the radiative decay widths $\Gamma(D^*\to D\gamma)$.
Currently, the latter are only available from the theory estimates. The LCSR prediction
can be found e.g. in \cite{Li:2020rcg,Rohrwild}.} and using the
precisely measured branching fraction ${\cal BR}(D^*\to D \pi)=0.677 \pm 0.005$ from \cite{PDG} yields
\be
\Gamma(D^{*+}\to D^0\pi^+)=
(56.5 \pm 1.3)\, {\rm keV} \,,
\ee
where we have added the two errors of independent measurements in quadrature.
Then, from (\ref{eq:strongD}) we finally obtain the strong coupling:
\be
[g_{D^*D\pi}]_{\rm exp}=16.8 \pm 0.2 \,.
\label{eq:DstDpiexp}
\ee
Our results on the $D^*D\pi$ coupling presented in Table~\ref{tab:res} are somewhat smaller than
the above value, but the difference is not significant.
Even if we take the smallest interval predicted from LCSR
(the combination of Model 2 with the lattice decay constants)
its upper limit is only 10\% smaller
than the measured strong coupling.

\begin{figure}[h]
\centering
\includegraphics[width=0.7 \textwidth]{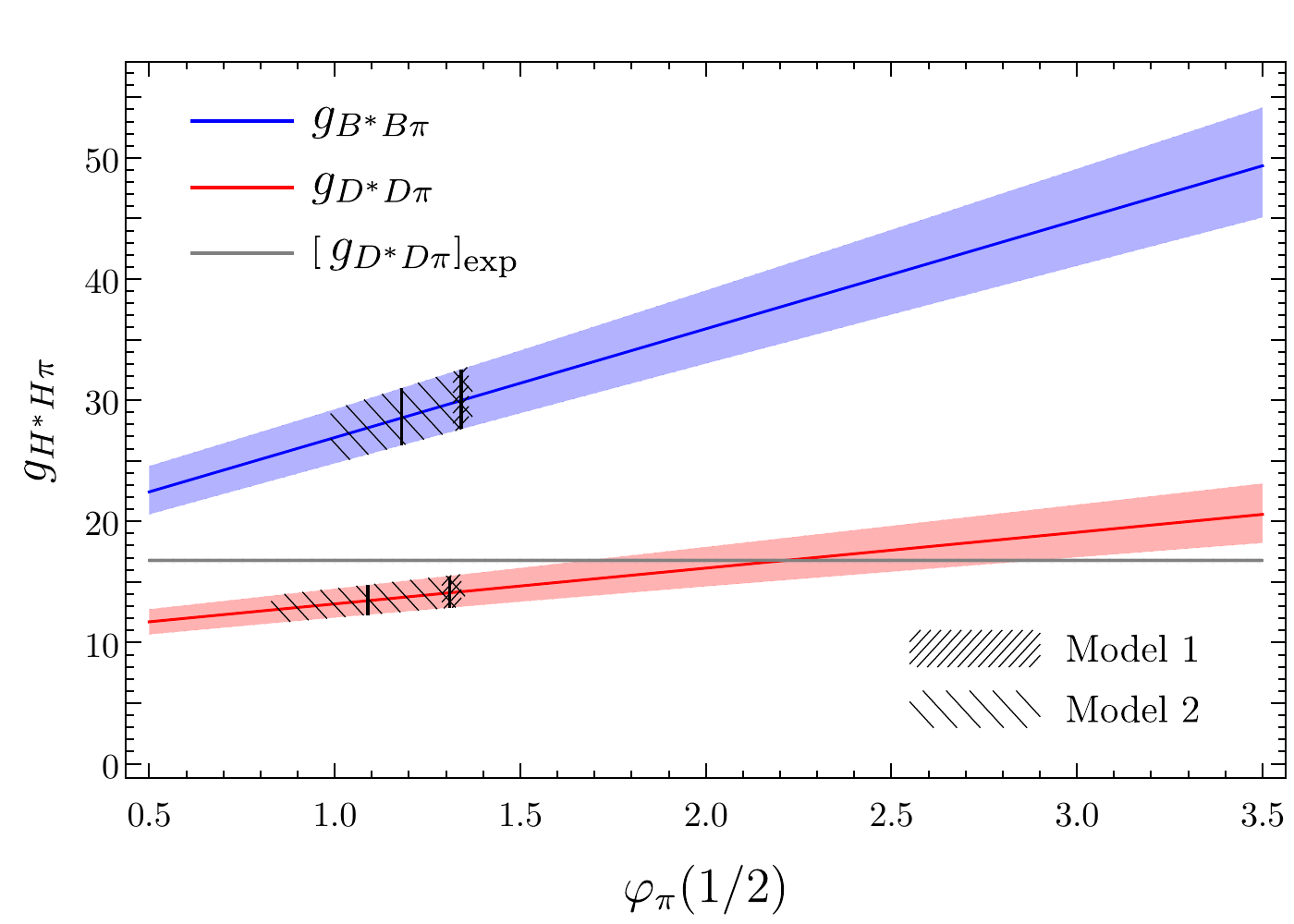}
\caption{The dependence of
$g_{D^*D\pi}$ and $g_{B^*B\pi}$ on the value of $\varphi_\pi(1/2)$.
The lattice QCD results for the decay constants of heavy mesons are employed.}
\label{fig:phi}
\end{figure}
Furthermore, it is  instructive to investigate how sensitive
are the  LCSRs for $D^*D\pi$ and $B^*B\pi$ couplings
to the midpoint value of the pion twist-2 DA.
In Figure \ref{fig:phi} we plot the dependence of both strong couplings
on $\varphi_\pi(1/2)$ considering the latter as a free parameter. We observe a very mild dependence of
$g_{D^*D\pi}$, so that
the overlap of the LCSR prediction within its uncertainty interval
with the experimental value (\ref{eq:DstDpiexp}) yields a broad interval with
$\varphi_\pi(1/2)>1.5$.
Having in mind that an unaccounted uncertainty of LCSR
at the level of $\sim 10\%$ is not excluded, we conclude that fixing the midpoint value of $\varphi_\pi$
only from the measured $D^*D\pi$ coupling is not realistic.
In this respect, the dependence of $g_{B^*B\pi}$ on $\varphi_\pi(1/2)$
plotted in Figure \ref{fig:phi} is steeper.
Hence, an accurate lattice QCD prediction for this coupling
obtained at finite $b$-quark mass,  being equated to LCSR, can yield
a more tight constraint on the pion DA.

Returning to the comparison of $g_{D^*D\pi}$ with experiment, a comment
is in order. The interval $\varphi_\pi(1/2,\mu=1.5 \,\GeV)>1.5$ preferred from this
comparison may indicate that the pion twist-2 DA at low scales has a structure
different from both  Models 1, 2 we have used. It is in fact possible to construct a pion DA which
has a midpoint value exceeding the asymptotic limit and,
simultaneously, the second Gegenbauer moment equal to the lattice QCD value
in (\ref{eq:a2latt}). The simplest option is to adopt a truncated Gegenbauer expansion
with a relatively large positive $a_4(1\, \GeV)$ and small $a_{n>4}$.
A pion DA with such a pattern of Gegenbauer moments at $\mu=1 \,\GeV$:
 $a_2=0.135\pm 0.032$ (the rescaled value (\ref{eq:a2latt})) and $a_4=0.218\pm 0.059$, $a_{n>4}=0$,
was used in \cite{Cheng:2020vwr} (see the second line in the Table IV there)
among other models fitting the LCSR for the pion electromagnetic form factor to the timelike form factor data.
According to (\ref{eq:phipi12}), the resulting midpoint value is
$\varphi_\pi(1/2,\, \mu=1 \,\mbox{GeV})= 1.81\pm 0.18$.
Note however that the Model 2 chosen above and adopted from the same analysis provides a better fit
to the pion form factor.

\begin{table}[h]
\centering
\begin{tabular}{|l||c|c|c|}
\hline
Method  &$g_{D^*D\pi}$  & $g_{B^*B\pi}$ & $\hat{g}$ \\[1mm]
\hline
&&&\\[-3mm]
LQCD, $N_f=2$ \; \cite{Becirevic:2012pf}& $15.9\pm 0.7^{+0.2}_{-0.4}$& --& --\\[1mm]
\hline
&&&\\[-3mm]
LQCD, $N_f=2+1$ \; \cite{Can:2012tx}&
$16.23\pm 1.71$ & --& --\\[1mm]
\hline
&&&\\[-3mm]
\multirow{2}{*}{LQCD, $N_f=2+1$  \; \cite{Flynn:2015xna}}
&\multirow{2}{*}{--}
&$\frac{2m_B}{f_\pi}(0.56\pm0.03\pm 0.07)$
&\multirow{2}{*}{--} \\[1mm]
&&$=45.3 \pm 6.0$&\\[1mm]
\hline
&&&\\[-3mm]
LQCD, $N_f=2$ \; \cite{Becirevic:2009yb} &-- & --&$0.44\pm 0.03^{+0.07}_{-0.0}$ \\[1mm]
\hline
&&&\\[-3mm]
LQCD, $N_f=2+1$ \; \cite{Detmold:2012ge}&-- & --&$0.449\pm 0.051$ \\[1mm]
\hline
&&&\\[-3mm]
LQCD, $N_f=2$ \; \cite{Bernardoni:2014kla}&-- &-- &$0.492\pm 0.029$ \\[1mm]
\hline
&&&\\[-2mm]
LCSR (this work) & $14.1^{+1.3}_{-1.2}$
& $30.0^{+2.6}_{-2.4}$
& $0.30^{+0.02}_{-0.02}$  \\[1mm]
\hline
\end{tabular}
\caption{Strong couplings of the heavy mesons with pion obtained from lattice QCD (LQCD),
compared with our LCSR prediction (using Model 1 of the pion DA and the lattice-QCD decay constants).}
\label{tab:compar}
\end{table}
In Table~\ref{tab:compar} we compare our results with
the strong couplings  calculated from the lattice QCD. We only select the results obtained with the number of flavours $N_f>1$.
The determinations of $g_{D^*D\pi}$ in \cite{Becirevic:2012pf,Can:2012tx} are consistent
with our results, whereas the only available result \cite{Flynn:2015xna} for $g_{B^*B\pi}$
is significantly larger than our prediction.
The same is valid for the static coupling $\hat{g}$ in the lattice QCD.
The latter  comparison is, however,  not completely consistent
because our results for $\hat{g}$ are based on the fit of Eq.~(\ref{eq:hqe})
which includes the higher-order correction in the heavy quark expansion.

Furthermore, let us mention an independent possibility to extract the $B^*B\pi$ coupling.
The procedure, explained in detail in \cite{Imsong:2014oqa}
(see also \cite{Li:2010rh}),
is based on the hadronic dispersion relation for
the $B\to \pi$ vector form factor:
\be
f^+_{B\pi}(q^2)=\frac{g_{B^*B\pi} f_{B^*}}{2m_{B^*}(1-q^2/m^2_{B^*})}+
\frac{1}{\pi}\!\!\!\!\!\!\int\limits_{(m_B+m_\pi)^2}^\infty \!\!\!\!\!dt\,
 \frac{\mbox{Im} f^+_{B\pi}(t)}{t-q^2}
\label{eq:Hpidisp}
\ee
valid without subtractions. The above relation contains
the vector-meson $B^*$  pole, which lies
slightly below the threshold \footnote{The separation of the $D^*$ pole
located above the $D\pi$  threshold is
not a straightforward  procedure, hence here we refrain  from
considering the same  method for the $D^*D\pi$ coupling.} $q^2=(m_B+ m_\pi)^2$.
A small width of this meson determined by the $B^*\to B\gamma$ decay
can safely be neglected in (\ref{eq:Hpidisp}).
The residue of the $B^*$ pole is a product of the $B^*B\pi$
coupling and the $B^*$ decay constant.
Multiplying both sides of (\ref{eq:Hpidisp})
by the denominator of the pole term, we take the limit
$q^2\to m_{B^*}^2$, removing the complicated
integral over hadronic spectral density, so that
\be
g_{B^*B\pi}= \frac{2m_{B^*}}{f_{B^*}}\lim_{q^2\to m_{B^*}^2}
\Big[\left(1-q^2/m^2_{B^*}\right)f^+_{B\pi}(q^2)\Big] \,.
\label{eq:BstarBpilimit}
\ee
Here we benefit from the fact that the pion mass is much smaller than $m_B$,
hence, any analytic representation of the expression
in the square bracket of the above equation,
valid in the low pion recoil region
$q^2\lesssim (m_B-m_\pi)^2$ of the $B\to\pi$ transition,
provides an accurate limit.
The most convenient representation for that purpose  is the
BCL version \cite{Bourrely:2008za} of the $z$-expansion
based on the conformal mapping of the momentum transfer squared:
$$ q^2\to z(q^2)= \frac{\sqrt{(m_B+m_\pi)^2-q^2}-\sqrt{(m_B+m_\pi)^2-t_0}}{
\sqrt{(m_B+m_\pi)^2-q^2}+\sqrt{(m_B+m_\pi)^2-t_0}}\,,$$
where the optimal choice is
$t_0= (m_B+m_\pi)(\sqrt{m_B}-\sqrt{m_\pi})^2$.
In this case,
the expression under the square bracket in (\ref{eq:BstarBpilimit})
represents a polynomial in $z$ and can be easily continued
to $z(m_{B^*}^2)$.

As an example, we make use of the lattice QCD, $N_f=2+1$ calculation \cite{Lattice:2015tia}
of the $B\to \pi$ form factor, where the $z$-expansion
\be
\left(1-q^2/m^2_{B^*}\right)f^+_{B\pi}(q^2)=\sum\limits_{n=0}^{N_z-1}b_n^+\Big( [z(q^2)]^n-
(-1)^{n-N_z}\frac{n}{N_z}[z(q^2)]^{N_z}\Big)\,,
\label{eq:Bpizexp}
\ee
was implemented.
Choosing the preferred
$N_z=4$ version with the coefficients $b_n^+$  from  Table XIV of \cite{Lattice:2015tia},
we substitute (\ref{eq:Bpizexp}) in (\ref{eq:BstarBpilimit}).
Adopting for $f_{B^*}$ the lattice QCD value in (\ref{eq:lattVP}), we obtain:
\be
g_{B^*B\pi}=34.5\pm 3.0 \,,
\label{eq:gBstpole}
\ee
where the uncertainty is obtained taking into account the errors and correlations
of the coefficients $b_n^+$ quoted in \cite{Lattice:2015tia} and the error
of the decay constant $f_{B^\ast}$.

Before commenting on this result, we note that
the truncated $z$-expansion in (\ref{eq:Bpizexp}) is employed
here merely as a smooth fit function. It is used to fit the l.h.s. of (\ref{eq:Bpizexp})
which consists of the form factor calculated \cite{Lattice:2015tia}  at $17\,\GeV^2 <q^2<26\,\GeV^2$
and multiplied with $(1-q^2/m^2_{B^*})$  to remove the $B^*$ pole .
We then extrapolate this function  to a slightly larger $q^2=m_{B^*}^2\simeq 28$ GeV$^2$
to reach the limit (\ref{eq:BstarBpilimit}).
Hence, the accuracy of the truncation in (\ref{eq:Bpizexp}) plays a minor role
in this procedure, being more important for an extrapolation to the small $q^2$ region.

The estimate (\ref{eq:gBstpole}) turns out to be significantly smaller than the
 result of a ``direct" lattice QCD calculation of $g_{B^*B\pi}$,
 involving the finite-mass $b$-quarks \cite{Flynn:2015xna}
and presented in Table~\ref{tab:compar}. On the other hand, the result in (\ref{eq:gBstpole})
is within uncertainty consistent with our LCSR result.
Finally, we also quote the earlier result
$g_{B^*B\pi}=30\pm 5$  obtained in \cite{Imsong:2014oqa}
using (\ref{eq:BstarBpilimit}) together with
the $z$-expansion of the $B\to\pi$ form factor
calculated from LCSR. Since this calculation was done at small and intermediate
momentum transfers, the extrapolation via $z$-expansion plays a more important role
and the estimated uncertainty is naturally larger than for the lattice QCD results.

\section {Conclusions and perspectives}
In this paper we revisited the calculation of the strong couplings $g_{D^\ast D\pi}$ and $g_{B^\ast B \pi}$
from the  LCSR that was originally derived in \cite{BBKR}.
The method is based on the OPE  of the
correlation function (\ref{eq:corr}) in terms of the pion DAs
with growing twist. The use of a finite-mass heavy quark
allows one to easily transform our sum-rule expressions  to be applied for both charmed and bottom mesons.

Our main new result is the  NLO twist-3 term in the LCSR,
calculated from the gluon radiative corrections to the underlying correlation function.
We also derived compact analytical expressions for
both twist-2 and twist-3 NLO terms  in a form of double dispersion relation.
This derivation was done in the  $\overline{\rm MS}$ scheme
for the heavy quark mass, and additional terms for a transition to the pole mass
scheme were also obtained for the sake of generality.
Among other new results, the continuum subtraction under the quark-hadron duality assumption is
extended to all twist-3 and twist-4 terms
at LO. We also carried out a detailed investigation of the sensitivity of LCSR to the form of two-dimensional duality region.
For the dominant part of the spectral density which is concentrated near the diagonal
on the plane of the two variables, the triangle region was found to be the most convenient
choice. In addition, all input parameters entering LCSR
were updated,
with an emphasis on the key parameter -- the midpoint value of the twist-2 pion DA.

As a result, the overall accuracy of the LCSRs for the strong couplings
is substantially improved with respect
to the earlier analyses in \cite{BBKR,KRWY_2}. Especially important, also numerically, is the inclusion of the new NLO twist-3 term in these sum rules.
Due to a more precise input, the parametrical
uncertainty for the default version of  LCSRs
-- with the heavy-meson
decay constants taken from the lattice QCD -- is reduced to the level of 10\%.
Still, not completely included in this estimate
is a ``systematic uncertainty"
caused by the quark-hadron duality which is more pronounced
for the double dispersion relation, than for the sum rules
based on the single-variable dispersion relation.
One step in assessing this uncertainty was done here by
examining the variation due to the shape of the duality region which was found to be  rather small.

We compared our result
for $g_{D^\ast D\pi}$ with the experimental measurement of this coupling inferred from
the $D^*\to D\pi$ decay branching fraction and the $D^*$ total width.
Identifying, somewhat qualitatively, as a $1 \, \sigma$ standard  deviation,
the estimated uncertainty of $O(10\%)$ ($O(20\%)$) of the LCSR result for $g_{D^\ast D\pi}$
when the lattice-QCD values (two-point  sum rules) are used for decay constants,
we find that our result is smaller by approximately $2 \, \sigma$ than
(agrees within $1 \, \sigma$ with) the measured value.
The agreement is much better than before, when only
the LO \cite{BBKR} or partial NLO \cite{KRWY_2} results were included in the numerical
analysis. We conclude that there is no need for a radical
modification of the quark-hadron duality ansatz, e.g. adding
explicitly the radially excited heavy-meson states to the
hadronic part of the sum rules, as suggested earlier \cite{Khodjamirian:2001bj,Becirevic:2002vp}.
 On the other hand, we found that the observed insignificant
deviation of the LCSR prediction for $g_{D^\ast D\pi}$
from experiment can be removed by a moderate modification
of the twist-2 pion DA at low scales.
This modification, in its turn, yields quite a noticeable
growth of our prediction for $g_{B^\ast B\pi}$.
Hence, the sum rule for strong coupling considered here
reveals a sensitivity to the pion twist-2 DA, similar to the other well-known LCSRs
for the $\gamma^{\ast} \gamma \to \pi$ transition form factor, the pion  e.m. form factor
and $ H\to \pi$ form factors ($H=D,B$).

Since there are no direct measurements of the $g_{B^\ast B\pi}$ coupling,
we can only compare our result with the lattice QCD calculations.
The most advanced calculation of \cite{Flynn:2015xna} performed at a finite $b$-quark mass,
however, neglecting the subleading power terms in the heavy quark expansion yields a $g_{B^\ast B\pi}$ coupling
which is about 30\% larger than the LCSR prediction (see Table~\ref{tab:compar}).
On the other hand, as we have demonstrated,
the latter is quite compatible with the value (\ref{eq:gBstpole})
extracted from the extrapolation to the $B^*$ pole of the lattice QCD
results for the $B\to \pi$ form factor \cite{Lattice:2015tia}.

Finally, one of the main conclusions following from our analysis
is that the inverse mass correction to the heavy-quark limit of the
strong couplings is quite large. Hence, it is probably premature to compare
our results for that limit with the effective coupling
$\hat{g}$ inferred from the lattice QCD calculations performed in HQET.
To clarify this issue, an alternative LCSR for the heavy-meson-pion strong coupling has to be derived in
HQET and compared with the heavy-mass expansion of the sum rule considered here.

The method of LCSR considered
in this paper is well suited to calculate a whole variety
of strong couplings involving the pion. This can easily be done
by varying the spin-parity or flavor quantum numbers
of both interpolating currents in the
correlation function. An early work in this direction
can be found e.g.
in \cite{Colangelo:1997rp} where the strong couplings of heavy mesons with
other combinations of spin-parities were calculated. Switching to the strange
quark in the interpolating currents allows one to access
the strange counterparts of the strong couplings
considered here, such as $g_{H_s^\ast H K}$ and $g_{H^* H_s K}$
for both  $H=D,B$.

The limited scope of this paper prevents us from discussing
in detail alternative  sum rules for the strong $H^*H\pi$ couplings, e.g.
the ones employing the external soft-pion field in which the correlation function of two heavy-light currents is expanded in terms of local operators
(see \cite{Eletsky:1984qs,Colangelo:1995ph,Grozin:1997qq}).
A detailed derivation of this method and comparison with LCSR can be found in \cite{BBKR}.
Due to the growing interest in the $B$-meson DAs, it will be
also interesting  to ``invert" the correlation function considered in this paper
to the one  with the $B$-meson DAs and the pion-interpolating current.

\section*{Acknowledgments}
First of all, we would like to thank Nils Offen and Patrick Gelhausen   for participating at the early stages of this work. The work of A.K. is supported by the DFG (German Research
Foundation) under grant  396021762-TRR 257
 ``Particle Physics Phenomenology after the Higgs Discovery''.
B.M. has been supported by the European Union through the European Regional Development Fund the Competitiveness and Cohesion Operational Programme (KK.01.1.1.06) and by the Croatian Science Foundation (HRZZ) project IP-2019-04-7094. B.M. would also like to express her gratitude to Wolfram Research for providing a free licence to use  {\fontfamily{cmtl}\selectfont Mathematica} at home during the COVID-19 pandemic lockdown and to Goran Duplan\v ci\'c for discussions.
A.K. and B.M. are also partially supported by the Alexander von Humboldt Foundation in the framework of the Research
Group Linkage Programme funded by the German Federal Ministry of Education and Research.
Y.M.W. acknowledges support from the National Youth Thousand Talents Program, the Youth Hundred Academic Leaders Program of Nankai University, the  National Natural Science Foundation of China  with Grant No. 11675082, 11735010 and 12075125,
and  the Natural Science Foundation of Tianjin with Grant No. 19JCJQJC61100.
Y.B.W. is supported in part by the Alexander-von-Humboldt Stiftung.
We also would like to express a special thanks to the Mainz Institute for Theoretical Physics (MITP) of the Cluster of Excellence PRISMA+ (Project ID 39083149) for its hospitality and support during the
scientific program "Light-cone distribution amplitudes in QCD" where this work was re-intiated.

\appendix
\section{Pion light-cone DAs}
\label{app:lcda}
The  definitions of the pion light-cone DAs in terms of the vacuum-to-pion matrix element
of the quark-antiquark and quark-antiquark-gluon nonlocal operators can be found in \cite{DKMMO}.
Here the expressions for pion DAs used in the numerical analysis are collected:
\begin{itemize}
\item  the twist-2 DA:
\vspace{-2mm}
\begin{align}
\varphi_\pi(u,\mu) = 6u\bar u  \left [ 1+\sum\limits_{n=2,4,..} a_n(\mu) C_n^{3/2}(u-\bar u) \right ] \,,
\label{eq:phipi2}
\end{align}
where $C_n^\alpha (z)$ are Gegenbauer polynomials
and the scale dependence of the moments reads
\begin{align}
\label{eq:rescal}
 a_n(\mu)= \left[\frac{\alpha_s(\mu)}{\alpha_s(\mu')}\right]^{\frac{\gamma^n_0}{\beta_0}} a_n(\mu'),
\end{align}
with $\beta_0=11-\frac{2}{3} n_f$ and the anomalous dimension \\
\begin{align}
\gamma^n_0=C_F \Big[-3-\frac{2}{(n+1)(n+2)}+4\sum\limits_{k=1}^{n+1}
\left(\frac{1}{k}\right) \Big] \,;
\end{align}
\end{itemize}

\begin{itemize}
\item the twist-3 two-particle DAs:
\vspace{-2mm}
\begin{align}
\label{eq:DAtw3}
\phi_{3\pi}^p(u)=1+30\frac{f_{3\pi}}{\mu_\pi f_\pi} C_2^{1/2}(u-\bar u)
-3 \frac{f_{3\pi}\omega_{3\pi}}{\mu_\pi f_\pi} C_4^{1/2}(u-\bar u)\,,\\
\phi_{3\pi}^\sigma(u)=6u\bar u \left [ 1
+5 \frac{f_{3\pi}}{\mu_\pi f_\pi}\Big(1-\frac{\omega_{3\pi}}{10}\Big)C_2^{3/2}(u-\bar u) \right ]\,;
\end{align}

\item the twist-3 three-particle DA:

\begin{align}
\Phi_{3\pi}(\alpha_i)= 360 \alpha_1\alpha_2\alpha_3^2 \left [1+\frac{\omega_{3\pi}}{2}(7\alpha_3-3) \right ] \,.
\end{align}
\vspace{-2mm}

Transforming the integration variables, the initial
expression for the contribution of this DA,
\begin{align}
F^{{\rm tw3},\bar qGq}(q^2,(p+q)^2)=4 m_Q f_{3\pi}\int_0^1 dv
\int \mathcal D\alpha \frac{v( q\cdot p) \Phi_{3\pi}(\alpha_i)}{(m_Q^2-(q+(\alpha_1+\alpha_3v)p)^2)^2} \,,
\end{align}
where $\mathcal{D}\alpha\equiv d\alpha_1d\alpha_2d\alpha_3 \, \delta(1-\alpha_1-\alpha_2-\alpha_3)$,
is reduced to (\ref{eq:OPEtw33})
and we use the following notation:
\begin{eqnarray}
\overline \Phi_{3\pi}(u) & \equiv & \frac{1}{2u}\int\limits_0^u d\alpha_1(u-\alpha_1)
\int\limits_{u-\alpha_1}^{1-\alpha_1}\frac{d\alpha_3}{\alpha_3^2}
\Phi_{3\pi}(\alpha_1,1-\alpha_1-\alpha_3,\alpha_3)
\nonumber\\
&=& 5 u^2\bar{u}^2 \, \left [3+\omega_{3\pi}\Big(-1+\frac74 u\Big) \right ] \,;
\end{eqnarray}

\item the twist-4  two-particle DAs:
\vspace{-2mm}
\begin{align}
 \psi_{4\pi}(u)=\frac{20}{3}\delta_\pi^2 C_2^{1/2}(2u-1)\,,
\end{align}

with a shorthand notation for the integral in (\ref{eq:OPEtw4psi})
\begin{align}
\bar \psi_{4\pi}(u) \equiv u \int^u_0 dv \, \psi_{4\pi}(v)=
\frac{20}{3}\delta_\pi^2 u^2\bar{u}(1-2u)\,,
\end{align}
and
\begin{eqnarray}
\phi_{4\pi}(u) &=&  \frac{200}{3} \delta^2_\pi u^2 \bar u^2 +
8 \delta^2_\pi \epsilon_\pi\Bigg [ u\bar u (2+13 u \bar u)
+ 2u^3(10-15 u +6 u^2) \ln(u)
\nonumber\\
&&  + 2\bar u^3 (10-15 \bar u + 6 \bar u^2) \ln(\bar u)\Bigg ] .
\end{eqnarray}
\end{itemize}

\begin{itemize}
\item the  twist-4 three-particle DAs:
\vspace{-2mm}
\begin{align}
 \Phi_{4\pi}(\alpha_i) &= 120 \delta^2_\pi \epsilon_\pi (\alpha_1 - \alpha_2)\alpha_1 \alpha_2 \alpha_3,\nonumber\\
 \Psi_{4\pi}(\alpha_i) &= 30 \delta^2_\pi(\alpha_1-\alpha_2)\alpha_3^2\Big(\frac{1}{3}+2\epsilon_\pi(1-2\alpha_3)\Big),\nonumber\\
 \tilde{\Phi}_{4\pi}(\alpha_i) &= -120 \delta^2_\pi \alpha_1 \alpha_2 \alpha_3 \Big(\frac{1}{3}+\epsilon_\pi(1-3\alpha_3)\Big),\nonumber\\
 \tilde{\Psi}_{4\pi}(\alpha_i) &= 30 \delta^2_\pi  \alpha_3^2 (1-\alpha_3)\Big(\frac{1}{3} +2\epsilon_\pi(1-2\alpha_3)\Big)\,.
\end{align}

Using these expressions, we transformed the sum of the contributions of the above DAs,
\begin{align}
F^{{\rm tw4},\bar qGq}(q^2,(p+q)^2)&= m_Q^2 f_\pi \int^1_0 dv \int \frac{\mathcal{D}\alpha}{(m_Q^2-(q+(\alpha_1+\alpha_3 v)p)^2)^2} \nonumber\\
&\Big [ 2\Psi_{4\pi}(\alpha_i)-\Phi_{4\pi}(\alpha_i) +2\tilde{\Psi}_{4\pi}(\alpha_i)-\tilde{\Phi}_{4\pi}(\alpha_i)\Big ] \Big|_{\alpha_2=1-\alpha_1-\alpha_3}\,,
\end{align}
to the compact form (\ref{eq:OPEtw43}), denoting
\begin{eqnarray}
\overline \Phi_{4\pi}(u) &=& \int\limits_0^u d\alpha_1(u-\alpha_1)
\int\limits_{u-\alpha_1}^{1-\alpha_1}\frac{d\alpha_3}{\alpha_3}
\Big [ 2\Psi_{4\pi}(\alpha_i)-\Phi_{4\pi}(\alpha_i)
+2\tilde{\Psi}_{4\pi}(\alpha_i)-\tilde{\Phi}_{4\pi}(\alpha_i)\Big ] \Big|_{\alpha_2=1-\alpha_1-\alpha_3}
\nonumber \\
&=& \frac{\delta^2_\pi}{3}u^3\bar{u} \left \{ 10 -5u\Big[1+3\epsilon_\pi\Big(1-\frac35 u \Big)\Big] \right \} \,.
\end{eqnarray}
\end{itemize}

\section{Double spectral density at NLO}
\label{app:nlo}

The expression for the twist-2 part of the NLO double spectral density
in the sum rule (\ref{eq:SR3}) calculated in the $\overline{\rm MS}$ scheme reads:
\begin{align}
&\hspace*{-0.7cm}~\rho^{\rm (tw2,NLO)}(r_1,r_2)
\nonumber \\
= &~ f_\pi \,\theta(r_2-1)\,\bigg\{6\,\bigg[(r_1-1)\,(r_2-1)\,
\bigg(
3\,\ln\frac{m^2_Q}{\mu^2} + \ln^2\frac{r_1-1}{r_2-1}
-\ln\frac{r_1-1}{r_2-1}\,\big(\ln\frac{r_1}{r_2}
-\frac{1}{r_2}+1\big)
\nonumber \\
& + \ln(r_2-1)\Big[ 4-\ln(r_1r_2)-\frac{1}{r_2}\Big]
-{\rm Li}_2(1-r_1) -3\,{\rm Li}_2(1-r_2)
+\frac{r_1}{r_2} -\frac{9}{2} -\frac{4}{3}\pi^2\bigg)
\nonumber \\
&+(2-r_1-r_2)\,\Big(3\,\ln\frac{m^2_Q}{\mu^2}-4\Big)
-r_2\,(r_1-1)\,\ln r_2 -\frac{3}{2}\,(r_2-1)^2
\bigg]\,\frac{d^2}{d{r_1}^2} \delta(r_1-r_2)
\nonumber \\
&+ 6\,\theta(r_1-1)\, (r_1-1)\,(r_2-1)\,\Big[
2\ln\frac{r_1-1}{r_2-1} -\ln\frac{r_1}{r_2} +\frac{1}{r_2} -1 \Big]
\, \frac{d^3}{d{r_1}^3} \ln|r_1-r_2|
\bigg\} \,,
\label{eq:tw2nlo}
\end{align}
where we use the rescaled variables
\begin{align}
r_1=\frac{s_1}{m^2_Q}\,,
\qquad {\rm and} \qquad
r_2=\frac{s_2}{m^2_Q}\,.
\end{align}
To convert this result to the pole-mass scheme for the heavy quark,
we are required to include the following term in the above expression
\begin{eqnarray}
\Delta \rho^{\rm (tw2,NLO)}_{\rm pole}(r_1,r_2)
&=& -6\, f_\pi\, \left (3\,\ln \frac{m^2_Q}{\mu^2} - 4 \right) \, \theta(r_2-1)  \,
\nonumber \\
&& \Big[
(r_1-1)\,(r_2-1)+2-r_1-r_2 \Big] \, \frac{d^2}{d{r_1}^2} \delta(r_1-r_2) \,,
\label{eq:addtw2}
\end{eqnarray}
with the replacement of  $m_Q\to m_{Q, \, \rm {pole}}$.

The desired twist-3 spectral density at NLO in $\alpha_s$ is calculated here for the first time:
\begin{align}
&~\hspace*{-0.8cm}\rho^{\rm (tw3,NLO)}(r_1,r_2)
\nonumber \\
= &~
f_\pi\,\frac{\mu_\pi}{m_Q}
\bigg\{\theta(r_2-1)\,\bigg[-2\,(r_1 + r_2)\,\Big(3\,\ln\frac{m^2_Q}{\mu^2}-4\Big)
+  4\,r_2\,(r_1-1) \,\ln\frac{r_2-1}{r_1-1}
\nonumber \\
&
+ \frac{2\,r_1}{r_2}\,(r_2-1)^2\,\ln(r_2-1)
- \Big[1 + r_2 + r_1\,(r_1+2)\,(r_2-1)\Big] \ln r_2
+ (r_1-1)\,(r_2-3)
\nonumber \\
&
+ 2\,(r_1 + r_2 - 2 r_1 r_2) \bigg(-\ln^2\frac{r_2-1}{r_1-1} -
    \ln(r_1-1) \ln\frac{r_2}{r_1} + 2 \ln(r_2-1)\ln r_2
    \nonumber \\
&  + \frac{1}{r_2}\,\ln\frac{r_2-1}{r_1-1} + 3\,{\rm Li}_2(1 - r_2) +
    {\rm Li}_2(1 - r_1) + \frac{4}{3}\,\pi^2 \bigg)\bigg] \frac{d^2}{d{r_1}^2}\,\delta(r_1-r_2)
\nonumber \\
& -2\, \theta(r_1-1)\,\theta(r_2-1)\,
\bigg[ (r_1+r_2-2\,r_1\,r_2)\,\Big(\ln\frac{r_2}{r_1}-2\,\ln\frac{r_2-1}{r_1-1}
 + \frac{1}{r_2}\Big)
 \nonumber \\
& +2\,r_2\,(r_1-1)\bigg]\,
\frac{d^3}{d{r_1}^3} \ln|r_1-r_2|
+4 \,\Big[\frac{d}{d r_1}\,\delta(r_1-1)\Big]\,
\delta(r_2-1)\,\Big(3\,\ln\frac{m^2_Q}{\mu^2}-4\Big)
\bigg\}\,.
\label{eq:tw3nlo}
\end{align}
The additional term needed to convert the above expression to the pole-mass scheme
is
\begin{align}
\Delta \rho^{\rm (tw3,NLO)}_{\rm pole}(r_1,r_2)
= &~\frac{f_\pi\,\mu_\pi}{m_Q} \,
\Big(3\,\ln\frac{m^2_Q}{\mu^2}-4\Big)\,
\bigg\{\theta(r_2-1)\,(3\,r_1 + 3\,r_2- 2\,r_1\,r_2)\,
\frac{d^2}{d{r_1}^2}\,\delta(r_1-r_2)
\nonumber \\
& -4\,\Big[\frac{d}{d r_1}\,\delta(r_1-1)\Big] \,
\delta(r_2-1)\bigg\} \,.
\end{align}

To facilitate the comparison of our results with those already derived in \cite{KRWY_2},
we introduce the two dimensionless variables
\barr
r = \frac{s_1 -m_Q^2}{s_2 - m_Q^2}\,, \quad \sigma = \frac{s_1}{m_Q^2} + \frac{s_2}{m_Q^2} -2\,.
\ea
Taking into account the fact that the double spectral densities will be integrated over
the variables $r$ and $\sigma$ (replacing $s_1$ and $s_2$) in the LCSR for the strong coupling $H^{\ast} H \pi$,
the method of integration by parts (IBP) can be further applied to the $r$-variable integral in order to
simplify the obtained lengthy expressions of (\ref{eq:tw2nlo}) and (\ref{eq:tw3nlo})
for the triangular duality region, which will result in the vanishing surface terms.
Under this circumstance, the twist-2 NLO spectral density (\ref{eq:tw2nlo})
can be cast into a more compact form
\begin{align}
&\hspace*{-0.4cm}\rho^{\rm (tw2,NLO)}(r,\sigma)
\nonumber \\
=&~  f_\pi\, \frac{(r+1)^2}{\sigma} \,
\Bigg \{ 3\, \delta(r-1) \,
\Big[\delta(\sigma)-\frac{1}{2}\,\theta(\sigma)\Big]\,
\Big( 3\,\ln\frac{m^2_Q}{\mu^2} -4 \Big)+
\theta(\sigma) \,\Bigg[\delta(r-1) \,
\Bigg(\,2 \,\pi^2
\nonumber \\
&
+  6 \,\text{Li}_2 \big(-\frac{\sigma}{2}\big)
+\ln \frac{\sigma}{2}\,
\Big[ \, 3\, \ln\frac{\sigma+2}{2}
-\frac{3 \,\sigma \,(\sigma+4)\,(3 \,\sigma+10)+72}{2\,(\sigma+2)^3} \Big]
+\frac{3}{4}\,\frac{\sigma\,(\sigma+12)+12}{(\sigma+2)^2}
\Bigg)
\nonumber \\
& + \theta(r)\,\frac{6 \,r}{(r+1)}\,
\Big( 2 \,\ln r + \ln\frac{1+ r +\sigma}{1+ r + r\, \sigma}
-\frac{\sigma}{1+r+\sigma} \Big )\,
\frac{d^3}{dr^3}\ln|r-1|\Bigg]\Bigg \} \,.
\label{eq:tw2rho}
\end{align}
\\
Switching to the pole-mass scheme of the heavy quark,
one needs to add the following expression
\barr
\Delta \rho^{\rm (tw2,NLO)}_{\rm pole}(r,\sigma)  = f_\pi\,
\frac{(r+1)^2}{\sigma} \,
\Big( 3\,\ln\frac{m^2_Q}{\mu^2} -4 \Big)
\,
\bigg\{3\,\delta(r-1)
\,\Big[\,\frac{1}{2}\,\theta(\sigma) \, - \delta(\sigma)\Big]
\bigg\} \,.
\label{eq:poeladd}
\ea
It is straightforward to verify that the sum of (\ref{eq:tw2rho}) and (\ref{eq:poeladd})
is equivalent to the expression obtained in \cite{KRWY_2} in the pole-mass scheme.
Along the same vein, we express the twist-3 double spectral density at ${\cal O}(\alpha_s)$
(\ref{eq:tw3nlo}) for the default duality region as follows
\begin{align}
&\hspace*{-0.4cm}\rho^{\rm (tw3,NLO)}(r,\sigma)
\nonumber \\
=&~  f_\pi
\,\frac{\mu_\pi}{m_Q}\, \frac{(r+1)^2}{\sigma} \,
\Bigg \{
\bigg(\delta (r-1) \,
\Big[\delta(\sigma)-2\,\delta'(\sigma)\Big]
+4\,(r+1) \, \delta(r)\,\delta'(\sigma)
\bigg)\,
\Big( 3\,\ln\frac{m^2_Q}{\mu^2} -4 \Big)
\nonumber \\
&
+\Big(\frac{4}{3}\,\pi^2 +1 \Big)
\delta (r-1)\, \delta(\sigma)
+\theta(\sigma) \,\Bigg[\delta (r-1) \Bigg ( \,
\frac{4}{3} \,\pi ^2
+4\, \text{Li}_2\big(-\frac{\sigma}{2}\big)
+\frac{1}{2} \,\Big(\frac{\sigma}{4}-\frac{4}{\sigma}\Big)
\ln\frac{\sigma+2}{2}
\nonumber \\
&
+\ln\frac{\sigma}{2}\, \Big(
\frac{\sigma^2+4}{2 \,(\sigma+2)^2}
+ 2\,\ln \frac{\sigma +2}{2}\Big)
+\frac{1}{2\,(\sigma +2)^2} \Big[
\frac{3\,\sigma\,\big[\sigma\,(\sigma+12)+80\big]}
{8 }
+ \frac{32}{\sigma}+62
\Big]
\Bigg )
\nonumber \\
&  +  \theta(r) \,\frac{2}{\sigma\, (r+1)}
\Bigg(\Big[ (1+r)^2 + 2\, r \,\sigma \Big] \,
\Big(2\, \ln r
+ \ln \frac{1+ r +\sigma}{1+ r + r \,\sigma}\Big)
- \frac{(1+r)^2 \, (r-1)}{1+r+\sigma} - 2 \,r \, \sigma \Bigg)
\nonumber \\
& \hspace{0.2 cm} \, \times \, \frac{d^3}{dr^3}\ln|r-1|
\Bigg]\Bigg \}\,,
\label{eq:tw3rho}
\end{align}
and
\begin{align}
\Delta \rho^{\rm (tw3,NLO)}_{\rm pole}(r,\sigma) =&~  f_\pi \,\frac{\mu_\pi}{m_Q}\,
\frac{(r+1)^2}{\sigma} \,
\Big( 3\,\ln\frac{m^2_Q}{\mu^2} -4 \Big)
\nonumber \\
&\times\bigg\{ \frac{1}{2}\, \delta(r-1)
\Big[\theta(\sigma)   + 2 \,\delta'(\sigma)-\,\delta(\sigma) \Big]
-4\,(r+1)\,\delta(r) \,\delta'(\sigma)
\bigg\} \,.
\end{align}

Finally, for completeness we present the pole-mass scheme
additions to the final NLO expressions in (\ref{eq:NLOBorel}), separately for the twist-2 and twist-3 parts:
\barr
\Delta f^{\rm (tw2)}_{\rm pole}(\sigma) &=& \Big( 3\,\ln\frac{m^2_Q}{\mu^2} -4 \Big)\,
3\,\Big[\,\frac{1}{2} - \delta(\sigma-0^+)\Big] \,,
\nonumber\\
\Delta f^{\rm (tw3)}_{\rm pole}(\sigma) &=& \Big( 3\,\ln\frac{m^2_Q}{\mu^2} -4 \Big)\,
\frac{1}{2} \Big[1 - \delta(\sigma-0^+)  -  4\,\delta'(\sigma-0^+)\, \Big]\,.
\label{eq:deltapole}
\ea

\section{Two-point sum rules for heavy meson decay constants at NLO}
\label{app:2pt}
Here the two-point QCD sum rules for the decay constants of the
heavy-light mesons $H$ and $H^*$ ($H=D,B$) are presented. Their expressions -- with the NLO, $O(\alpha_s)$
accuracy in the perturbative and quark-condensate terms and including the dimension $d\leq 6$ condensates
-- are taken from \cite{GKPR}.
In these sum rules, we denote the Borel parameter squared and duality thresholds by the barred quantities
$\bar{M}^2$ and $\bar{s}_0,\bar{s}_0^*$, to distinguish them from the
analogous parameters in the  LCSR for the strong coupling.

For the decay constant of the pseudoscalar meson we have:
\ba
f_{H}^2=\frac{e^{m_{H}^2/\bar{M}^2}}{m_{H}^4 }
\Bigg\{\int\limits_{m_Q^2}^{\bar{s}_0}\!\!ds \, e^{-s/\bar{M}^2}\rho^{\rm (pert)}_{5}(s)
+\Pi_{5}^{\langle \bar{q} q \rangle} (\bar{M}^2)+
\Pi_{5}^{(d456)} (\bar{M}^2)
\Bigg \}\,,
\label{eq:fBSR}
\ea
where the perturbative spectral density (for a massless light quark) is:
\begin{align}
\rho_{5}^{(\rm pert)}(s)&=
\frac{3m_Q^2}{8\pi^2}s\left(1-z\right)^2
+\frac{3\alpha_s C_F}{16\pi^3}m_Q^2 s (1-z)
\Bigg[\frac{9}{2}(1-z)
\nonumber\\
&+2(1-z)
\Big[2\,{\rm{Li}}_2(z)+\ln z\ln(1-z)\Big]
+(3-z)(1-2z)\ln z
\nonumber\\
&-(1-z)(5-2z)\ln(1-z)
+(1-3z)\left(3\ln\frac{\mu^2}{m_Q^2}+4\right)\Bigg]\,,
\label{eq:pertNLO5}
\end{align}
denoting $z=m_Q^2/s$. The quark condensate contribution is
\begin{align}
 \Pi^{\langle \bar q q\rangle}_5 (\bar{M}^2)&= -m_Q^3\langle \bar q q\rangle e^{-\frac{m_Q^2}{\bar{M}^2}}
 \nonumber\\
 &\times\Bigg \{ 1-\frac{\alpha_sC_F}{2\pi}\Big[\Big(3\ln\frac{\mu^2}{m_Q^2}+4\Big)\frac{m_Q^2}{\bar{M}^2}-7-3\ln\frac{\mu^2}{m_Q^2}
 + 3 \, \Gamma\Big(0,\frac{m_Q^2}{\bar{M}^2}\Big)e^{\frac{m_Q^2}{\bar{M}^2}}\Big]\Bigg \} \;,
\end{align}
containing the incomplete gamma function $\Gamma(a,z) = \int_z^\infty t^{a-1} e^{-t} dt$.

The sum of contributions of the $d=4,5,6$ (gluon, quark-gluon and four-quark) condensates is
\begin{align}
\Pi_{5}^{(d456)} (\bar{M}^2)&= \Bigg[\frac{ \langle  GG\rangle m_Q^2}{12}
 -\frac{m^2_0\langle \bar q q\rangle m_Q^3}{2\bar{M}^2}
\Big(1-\frac{m_Q^2}{2\bar{M}^2}\Big)
\nonumber\\
& \hspace{0.5 cm} -\frac{16\pi r_{vac}\alpha_s\langle \bar{q} q \rangle^2m_Q^2}{27 \bar{M}^2} \Big(1-\frac{m_Q^2}{4\bar{M}^2}-\frac{m_Q^4}{12\bar{M}^4}\Big)
\Bigg] e^{-\frac{m_Q^2}{\bar{M}^2}}~\,.
\end{align}
In the above equations, the following shorthand notations
and parameterizations are used for
the QCD vacuum condensate densities:
\begin{align}
\langle 0|\bar{q} q |0\rangle\equiv
\langle \bar{q} q \rangle\,,
\qquad
(\alpha_s/\pi)\langle 0| G^a_{\mu\nu} G^{a\,\mu\nu} |0\rangle \equiv\langle GG\rangle  \,,
\qquad
\langle 0 |g_s\bar{q} G^a_{\mu\nu}t^a \sigma^{\mu\nu}q|0 \rangle = m_0^2 \langle \bar{q}q\rangle\,,
\nonumber
\end{align}
and the four-quark condensate density, factorized \cite{Shifman:1978bx} via vacuum insertion, is approximated by the square of quark condensates $r_{vac}\langle \bar{q}q\rangle ^2$, where
the numerical factor $r_{vac}$
parameterizes deviation from the factorization.
 Note that $m_0^2$ and the four-quark density
 multiplied by $\alpha_s$
 are to a good precision scale-independent. The numerical values of all condensate parameters
are presented in Table~\ref{tab:QCD}.

The sum rule for the decay constant of the vector heavy-light meson is:
\ba
f_{H^*}^2= \frac{e^{m_{H^*}^2/\bar{M}^2}}{m_{H^*}^2 }\Bigg [ \,
\int\limits_{m_Q^2}^{\bar{s}^*_0}\!\!ds \, e^{-s/\bar{M}^2}\rho^{\rm (pert)}_{T}(s)
+\Pi_{T}^{\langle \bar{q} q \rangle} (M^2)+\Pi_{T}^{(d456)} (M^2) \Bigg ] \,,
\label{eq:fBstarSR}
\ea
with the perturbative spectral density
\begin{align}
\rho_{T}^{(\rm pert)}(s)&=\frac{1}{8\pi^2}s\left(1-z\right)^2\left(2 +z\right)
+\frac{3\alpha_s C_F}{16\pi^3}s
\Bigg[1-\frac{5}{2}z+\frac{2}{3}z^2+\frac{5}{6}z^3
\nonumber\\
&+\frac{2}{3}(1-z)^2(2+z)\Big[2\,{\rm{Li}}_2(z)
+\ln z \ln(1-z)\Big]
+\frac{1}{3}z(-5-4z+5z^2)\ln z
\nonumber\\
&-\frac{1}{3}(1-z)^2(4+5z)\ln(1-z)
-z(1-z^2)\left(3\ln\frac{\mu^2}{m_Q^2}+4\right)\Bigg]\,,
\label{eq:pertNLOT}
\end{align}
and the condensate contributions:
\begin{align}
 \Pi^{\langle \bar q q\rangle}_T (\bar{M}^2)&=-m_Q\langle \bar q q\rangle e^{-\frac{m_Q^2}{\bar{M}^2}} \left \{ 1+\frac{\alpha_sC_F}{2\pi}\left[1-3\frac{m_Q^2}{\bar{M}^2}\ln\frac{\mu^2}{m_Q^2}-4\frac{m_Q^2}{\bar{M}^2}\right.\right.\\\nonumber
 &\left.\left.+\frac{m_Q^2}{\bar{M}^2}e^{\frac{m_Q^2}{\bar{M}^2}}\Gamma\Big(-1,\frac{m_Q^2}{M^2}\Big)\right]\right \} \;,
\end{align}

\begin{align}
\Pi_{T}^{(d456)} (\bar{M}^2) &=
\Bigg[- \frac{\langle GG \rangle }{12}
+\frac{ m^2_0\langle \bar q q \rangle m_Q^3}{4\bar{M}^4}
-\frac{32\pi\alpha_sr_{vac}\langle\bar{q} q\rangle^2}{81\bar{M}^2} \Big(1+\frac{m_Q^2}{\bar{M}^2}-\frac{m_Q^4}{8\bar{M}^4}\Big)\Bigg] e^{-\frac{m_Q^2}{\bar{M}^2}}\,.
\end{align}


\end{document}